\documentclass[numberedappendix]{emulateapj}
\usepackage{amsmath}

\slugcomment{ApJ, submitted; \today}

\newcommand{\G}{\Gamma}

\newcommand{\sT}{\sigma_{\rm T}}
\newcommand{\p}{^\prime}
\newcommand{\e}{\epsilon}
\newcommand{\ebar}{\bar{\e}}
\newcommand{\g}{\gamma}
\newcommand{\gp}{\gamma^{\prime}}

\newcommand{\ep}{\epsilon^\prime}

\newcommand{\dD}{\delta_{\rm D}}
\newcommand{\up}{u^\prime}

\newcommand{\psim}{\lower.5ex\hbox{$\; \buildrel \propto \over\sim \;$}}
\newcommand{\lbar}{\lower.0ex\hbox{$\; \buildrel
{\lower0.0ex \hbox{-}} \over\lambda  \;$}}

\newcommand{\tilr}{\tilde{r}}
\newcommand{\tilR}{\tilde{R}}
\newcommand{\tilx}{\tilde{x}}

\newcommand{\tils}{\tilde{s}}
\newcommand{\tilell}{\tilde{\ell}}

\newcommand{\meanesangle}{\langle\e_s\rangle_{\mu,\phi}}

\newcommand{\Li}{\mathrm{Li}}

\newcommand{\cm}{\mathrm{cm}}
\newcommand{\km}{\mathrm{km}}
\newcommand{\erg}{\mathrm{erg}}

\newcommand{\GeV}{\mathrm{GeV}}

\newcommand{\s}{\mathrm{s}}
\newcommand{\dday}{\mathrm{day}}

\newcommand{\hour}{\mathrm{hour}}

\newcommand{\Hz}{\mathrm{Hz}}
\newcommand{\pc}{\mathrm{pc}}

\newcommand{\Kelvin}{\mathrm{K}}

\shorttitle{External Compton Scattering in Blazar Jets}
\shortauthors{Finke}

\begin{document}
\title{External Compton Scattering in Blazar Jets and the Location of
the Gamma-Ray Emitting Region}

\author{Justin D.\ Finke}

\affil{U.S.\ Naval Research Laboratory, Code 7653, 4555 Overlook Ave.\ SW,
        Washington, DC,
        20375-5352\\
}

\email{justin.finke@nrl.navy.mil}

\begin{abstract}

I study the location of the $\gamma$-ray emission in
blazar jets by creating a Compton-scattering approximation valid for
all anisotropic radiation fields in the Thomson through Klein-Nishina
regimes, which is highly accurate and can speed up numerical
calculations by up to a factor $\sim10$.  I apply this approximation
to synchrotron self-Compton, and external Compton-scattering of
photons from the accretion disk, broad-line region (BLR), and dust
torus.  I use a stratified BLR model and include detailed
Compton-scattering calculations of a spherical and flattened BLR.  I
create two dust torus models, one where the torus is an annulus, and
one where it is an extended disk.  I present detailed calculations of
the photoabsorption optical depth using my detailed BLR and dust torus
models, including the full angle dependence.  I apply these
calculations to the emission from a relativistically moving blob
traveling through these radiation fields.  The ratio of $\gamma$-ray
to optical flux produces a predictable pattern that could help locate
the $\gamma$-ray emission region.  I show that the bright flare from
3C 454.3 in 2010 November detected by the Fermi Large Area Telescope
is unlikely to originate from a single blob inside the BLR since it
moves outside the BLR in a time shorter than the flare duration,
although emission by multiple blobs inside the BLR is possible; and
$\gamma$-rays are unlikely to originate from outside the BLR from
scattering of photons from an extended dust torus, since then the
cooling timescale would be too long to explain the observed short
variability. 

\end{abstract}


\section{Introduction}
\label{intro}

Blazars are active galactic nuclei (AGN) with jets of nonthermal
plasma moving at relativistic speeds oriented close to our line of
sight.  They produce nonthermal radiation across the electro-magnetic
spectrum, from radio to $\g$ rays, the lower portion of which is
almost certainly produced by nonthermal synchrotron.  Blazars are
classified based on the strength or weakness of broad lines in their
optical spectra as FSRQs or BL Lacs, respectively
\citep[e.g.][]{marcha96,landt04}, and on the location of their
synchrotron peak \citep[$\nu_{pk}$;][]{abdo10_sed} in $\nu F_{\nu}$
spectral energy distributions (SEDs) as low-synchrotron peaked (LSP;
$\nu_{pk}<10^{14}\ \Hz$), intermediate-synchrotron peaked (ISP;
$10^{14}\ \Hz < \nu_{pk} < 10^{15}\ \Hz$), or high-synchrotron peaked
(HSP; $10^{15}\ \Hz\ < \nu_{pk}$).

In the 0.1--300 GeV energy range, blazars dominate the sky in terms of
number of associated sources \citep{acero15_3fgl}.  Hadronic models
for $\g$-ray production in these sources often face difficulties
due to requiring excessive jet powers, especially for FSRQs and LSPs
\citep{boett13,zdziarski15,petro15_3c279,petro16}, although there are
some examples, especially HSPs, where jet powers are more reasonable
\citep[e.g.,][]{cerruti15}.  Inverse Compton scattering of soft
photons by relativistic nonthermal electrons in the jet is
usually invoked as the most likely mechanism for $\g$-ray production
in FSRQs and some LSPs.  Possible seed photons for Compton
scattering include the synchrotron photons produced by the same
population of electrons that produces the $\g$ rays \citep[known as
synchrotron self-Compton, or SSC;][]{bloom96}; or seed photons from
another portion of the jet
\citep[e.g.,][]{ghisellini05,macdonald15,sikora16}; or by seed photons
external to the jet entirely, for instance, from the accretion disk
\citep{dermer92,dermer93}, broad line region \citep[BLR;][]{sikora94},
or dust torus \citep{kataoka99,blazejowski00} in the object; or from
the cosmic microwave background
\citep[CMB;][]{boett08,yan12_1101,meyer15,sanchez15,zacharias16}.
The dominant seed photon source depends critically on the
location in the jet of the emitting region.  In order of increasing
distance from the black hole, the dominant external seed photon source
could be the accretion disk, the BLR, the dust torus, and/or
the CMB.  The dominant seed photon source, and the location of the
primary emitting region, is an important topic in the understanding of
blazar jets that has not yet been resolved.  I endeavor to make
progress in answering this question by providing detailed calculations
of Compton scattering of the relevant external radiation fields.

Calculations of Compton scattering of various external radiation
fields (external Compton or EC hereafter) using the full Compton cross
section, valid in the Thomson and extreme Klein-Nishina regimes,
including anisotropic external radiation fields, have been explored by
many authors
\citep[e.g.,][]{boett97,boett00,dermer09,hutter11,hunger16}.  These
calculations can be quite numerically intense, especially if
calculations are repeated numerous times in, for example, fitting
routines \citep{finke08_SSC,mank10,mank11,yan13,cerruti13_fit} or
multi-zone models \citep[e.g.,][]{jamil12,joshi14}.  Often,
$\delta$-function approximations, valid in the Thomson regime, are
used to approximate Compton scattering processes
\citep[e.g.,][]{dermer92,dermer93}.  In these approximations, {\em
all} of the scattering photons are assumed to have the same energy,
that of the mean scattered photon energy.  A $\delta$-function
approximation valid at all energies, in the Thomson through extreme
Klein-Nishina regimes, was developed by \citet{moderski05} for
isotropic radiation fields, which I made use of recently to compute
theoretical power spectral densities and Fourier-frequency dependent
time lags of blazar light curves \citep{finke15}.  In the present
manuscript, I generalize the $\delta$-function approximation of
\citet{moderski05} to anisotropic radiation fields (Section
\ref{deltafcnsection}) and apply it to EC for several external
isotropic and anisotropic external radiation fields (Section
\ref{ECsection}), and, for completeness, to SSC (Section
\ref{SSCsection}), all in the context of relativistic jets.
Researchers may find this $\delta$-function approximation useful for
Compton-scattering calculations in other astrophysical contexts
besides blazars, such as microquasars
\citep[e.g.,][]{gupta06a,dubus08,dubus10,zdziarski14}, colliding winds
of massive stars \citep[e.g.,][]{reimer06_wind}, or gamma-ray bursts
\citep[e.g.,][]{lu15}.  The speed and accuracy of the approximations
are explored in Section \ref{numericalsection}.  The $\delta$-function
approximations are used to derive the beaming pattern for the
scattering of various external radiation fields by relativistic jets
in Section \ref{beampatternsection}.

Another process necessary to the Compton-scattering model for blazar
jet emission is $\g\g$ absorption.  The interaction of $\g$ rays with
soft photons from the accretion disk, BLR, and dust torus can limit
the escape of $\g$ rays.  This process is explored in Section
\ref{absorbsection}.  As the $\g$-ray emitting region moves at
relativistic speeds, it travels through regions where various external
radiation fields dominate the Compton scattering process.  In Section
\ref{signaturesection} I look at the effect this would have on the
ratio of $\g$-ray to optical flux that one would expect as a function
of time.  This can give critical clues to the location of the
$\g$-ray emitting region in blazars.  In particular, I apply the
calculations presented here to the optical and $\g$-ray light curves
of the giant flare in 3C 454.3 in 2010 November.  I conclude with a
summary in Section \ref{discussion}.  In Appendix \ref{BLRmodel}, I
present a simple model for determining the luminosities and radii of
line emission in a stratified BLR based on a composite quasar spectrum
from the Sloan Digital Sky Survey (SDSS).

\section{Delta Function Scattering Approximation}
\label{deltafcnsection}

\subsection{Compton Scattering Cross Section in the Head-On Approximation}

The emissivity from Compton scattering is given by
\begin{flalign}
\label{emissivity1}
\e_s\, j(\e_s,\Omega_s) & = m_ec^3\e_s^2 \int d\Omega \int d\e\ n_{ph}(\e,\Omega)
\nonumber \\ & \times
\int d\Omega_e \int d\g\ n_e(\g, \Omega_e)
\nonumber \\ & \times
(1-\cos\psi) \frac{d\sigma}{d\e_sd\Omega_s}\ 
\end{flalign}
\citep[e.g.,][]{dermer09_book} where $m_e$ is the electron mass, $c$
is the speed of light, $\e_s$ is the scattered photon energy in
$m_ec^2$ units, $n_{ph}(\e,\Omega)$ is the number density of incident
(``seed'') photons with energies between $\e$ and $\e+d\e$ and solid
angles between $\Omega$ and $\Omega+d\Omega$, $n_e(\g,\Omega_e)$ is
the number density of incident electrons with Lorentz factors between
$\g$ and $\g+d\g$ and solid angles between $\Omega_e$ and
$\Omega_e+d\Omega_e$, $d\sigma/(d\e_sd\Omega_s)$ is the differential
cross section per scattered photon energy and solid angle, and $\psi$
is the angle between the direction of the incident photon and
electron.  

\begin{figure}
\vspace{2.2mm} 
\epsscale{0.8} 
\plotone{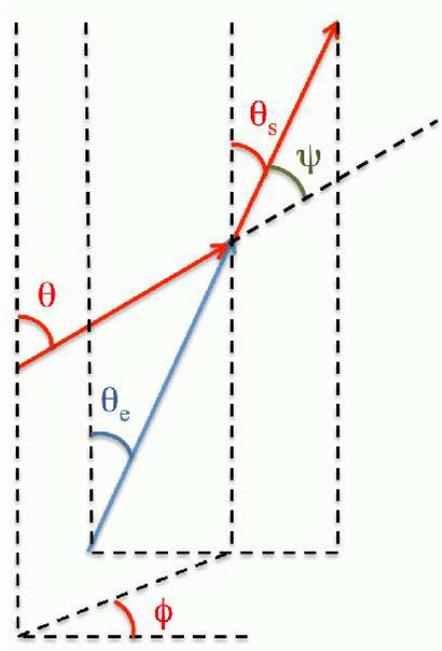}
\caption{An illustration of the geometry of scattering in the head-on approximation.}
\label{geometry_scatter}
\vspace{2.2mm}
\end{figure}

In the ``head-on'' approximation, $\e \ll 1 \ll \g$, and photons are
scattered in approximately the same direction as the incident
electrons \citep{reynolds82,dermer93,dermer09}, so that
\begin{flalign}
\frac{d\sigma}{d\e_sd\Omega_s}\ \approx \frac{d\sigma}{d\e_s}\delta(\Omega_s-\Omega_e)\ 
\end{flalign}
where
\begin{flalign}
\frac{d\sigma}{d\e_s} & = \int d\Omega_s \frac{d\sigma}{d\e_sd\Omega_s} = 
\frac{3\sT}{8\g\ebar}\ \Xi(\ebar,y)\ 
\nonumber \\ & \times 
H\left(\e_s; \frac{\ebar}{2\g}, \frac{2\g\ebar}{1+2\ebar}\right)\ ,
\end{flalign}
\begin{flalign}
\Xi(\ebar,y) & = 1-y + \frac{1}{1-y} - 
\nonumber \\ &
\frac{2y}{\ebar(1-y)} + \left[\frac{y}{\ebar(1-y)}\right]^2\ ,
\end{flalign}
\begin{flalign}
\ebar = \g\e(1-\cos\psi)\ ,
\end{flalign}
\begin{flalign}
y = \frac{\e_s}{\g}\ ,
\end{flalign}
and
\begin{equation}
H(x; a, b) = \left\{ \begin{array}{ll}
1 & a < x <b \\
 0 & \mathrm{otherwise}
\end{array}
\right. \ .
\end{equation}
If the incident photon travels in a direction given by
azimuthal and polar angles $\phi$ and $\theta=\arccos(\mu)$,
respectively, and the scattered photon travels in a direction given by
the azimuthal and polar angles, $\phi_s$ and
$\theta_s=\arccos(\mu_s)$, respectively, then
\begin{flalign}
\cos\psi & = \mu\mu_s + \sqrt{1-\mu^2}\sqrt{1-\mu_s^2}
\nonumber \\ & \times 
\cos(\phi-\phi_s)\ .
\end{flalign}
For the remainder of this paper, I define the coordinate system so
that $\phi_s=0$.  The geometry of the head-on approximation is
illustrated in Figure \ref{geometry_scatter}.  The head-on approximation
implies 
\begin{flalign}
\label{emissivity2}
\e_s\, j(\e_s,\Omega_s) & \approx m_ec^3\e_s^2 \int d\Omega \int d\e\ n_{ph}(\e,\Omega)
\nonumber \\ & \times
\int d\g\ n_e(\g, \Omega_s)(1-\cos\psi) \frac{d\sigma}{d\e_s}\ .
\end{flalign}

\begin{figure}
\vspace{10.0mm} 
\epsscale{1.0} 
\plotone{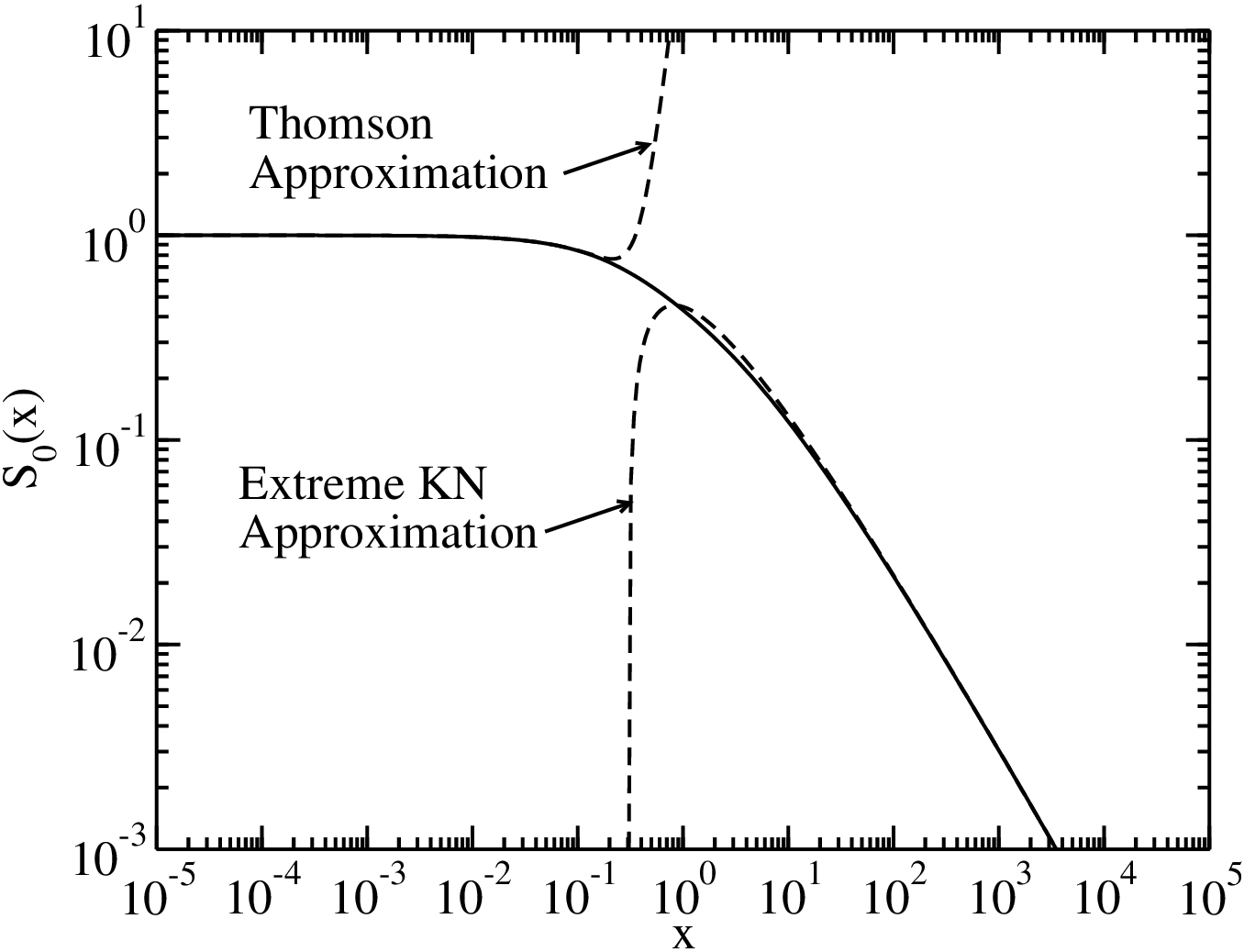}
\caption{The function $S_0(x)$ given by Equation (\ref{S0eqn}) and the approximations 
from Equations (\ref{S0Thom}) ($x\ll 1$) and (\ref{S0KN}) ($x\gg 1$).  }
\label{S0plot}
\vspace{2.2mm}
\end{figure}

It will be useful to compute the total Compton-scattering cross
section, which is
\begin{flalign}
\sigma(\ebar) = \int d\e_s \frac{d\sigma}{d\e_s} = 
\sT S_0(\ebar)
\end{flalign}
where
\begin{flalign}
\label{S0eqn}
S_0(x) & = \frac{3}{8x^2} \Biggr[ 4 + \frac{2x^2(1+x)}{(1+2x)^2} 
\nonumber \\ &
+ \frac{x^2-2x-2}{x}\ln(1+2x) \Biggr]\ .
\end{flalign}
This function has the asymptotes
\begin{flalign}
\label{S0Thom}
S_0(x) & \rightarrow 1 - 2x + \frac{26x^2}{5} - \frac{133x^3}{10} + 
\frac{1444x^4}{35}
\end{flalign}
for $x \ll 1$ and
\begin{flalign}
\label{S0KN}
S_0(x) \rightarrow \frac{3}{8x}\left[ \ln(x) + \frac{1}{2} + \frac{1}{2}\ln(4)\right]
\end{flalign}
for $x \gg 1$,
and it is plotted in Figure \ref{S0plot}.

\begin{figure}
\vspace{10.0mm} 
\epsscale{1.0} 
\plotone{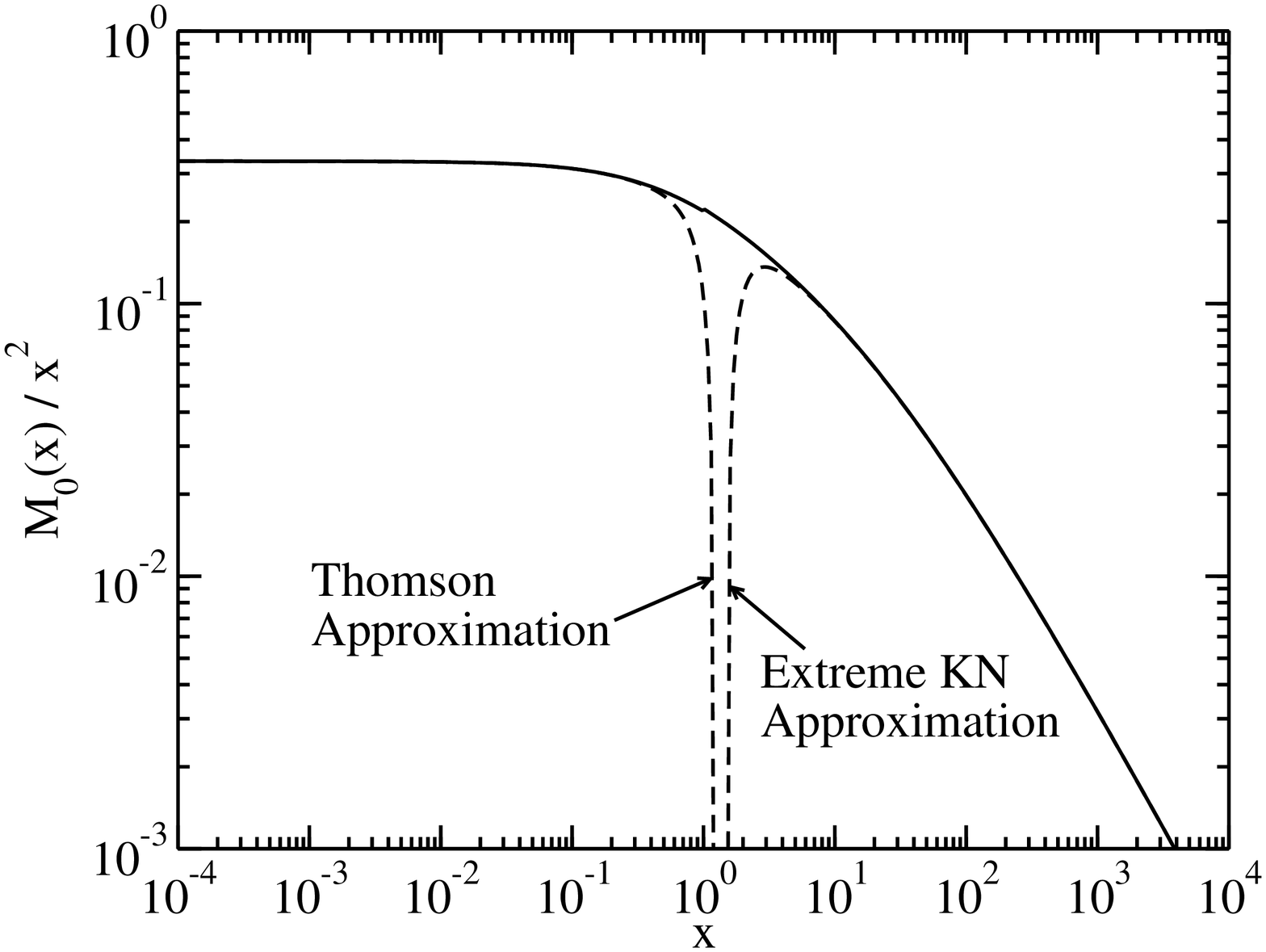}
\caption{The function $M_0(x)$ given by Equation (\ref{M0_1}) and the approximations
from Equations (\ref{M0Thom}) ($x\ll 1$) and (\ref{M0KN}) ($x\gg 1$).  }
\label{M0plot}
\vspace{2.2mm}
\end{figure}

It will also be useful to integrate the total cross section over all
angles, so that
\begin{flalign}
\int_0^{2\pi} d\phi \int_{-1}^{1} d\mu\ (1-\mu)\ \sigma(\g\e(1-\mu)) 
\nonumber \\ 
= \frac{3\sT(2\pi)}{8(\g\e)^2} M_0(4\g\e)
\end{flalign}
where
\begin{flalign}
\label{M0_1}
M_0(x) = 2(x+1)x\Li_2(-x) - \frac{1}{2}\left(\frac{x}{2}+1\right)x^2 
\nonumber \\
+ \left(\frac{x}{2}+4\right)(x+1)^2\ln(x+1) - 4\ 
\end{flalign} 
and $\Li_2(x)$ is the dilogarithm function.  The function $M_0(x)$ has
the asymptotes
\begin{equation}
\label{M0Thom}
M_0(x) \rightarrow \frac{x^2}{3} - \frac{2x^3}{9} + \frac{13x^4}{60} - \frac{133x^5}{600} 
\end{equation} 
for $x \ll 1$ and
\begin{flalign}
\label{M0KN}
M_0(x) & \rightarrow \frac{x}{2}\left(\ln(x)-\frac{1}{2}\right) + 3.114\ln(x) 
\nonumber \\
& - \ln^2\left(\frac{x}{2}\right) - 6.559 + \frac{2}{x}(2\ln(x) + 3.249) 
\end{flalign} 
for $x \gg 1$,
and it is plotted in Figure \ref{M0plot}.

\subsection{Mean Scattered Photon Energy}
\label{meansection}

\begin{figure}
\vspace{10.0mm} 
\epsscale{1.0} 
\plotone{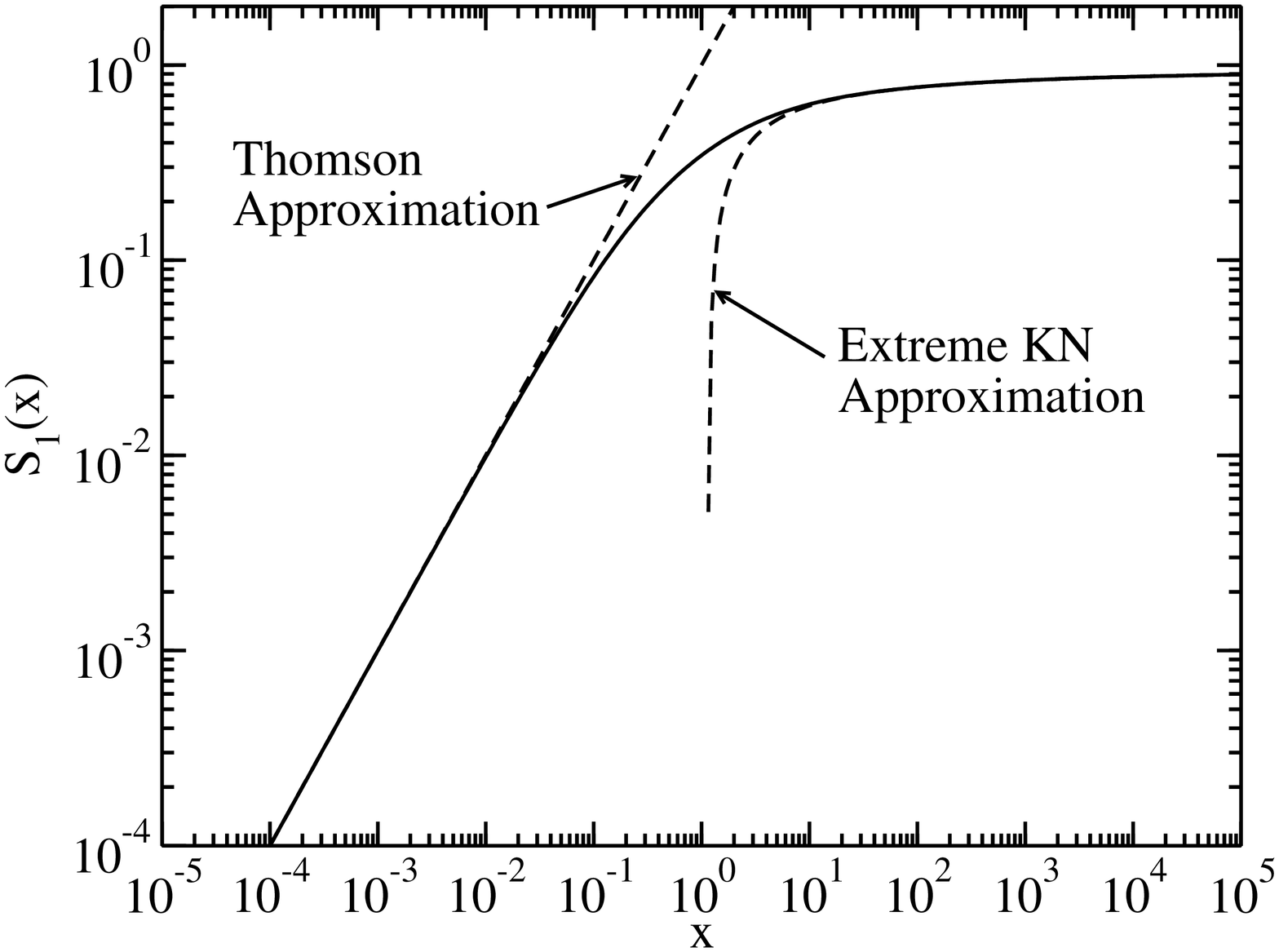}
\caption{The function $S_1(x)$ defined by Equation (\ref{S1}) and approximations from
Equation (\ref{S1asympt}) for $x\ll 1$ and $x\gg 1$.
}
\label{S1plot}
\vspace{2.2mm}
\end{figure}

The zeroth moment of the Compton-scattering cross section is
\begin{flalign}
\langle \epsilon_s^0\sigma \rangle & = 
\int d\Omega_s \int d\e_s \frac{d\sigma}{d\e_sd\Omega_s} 
\nonumber \\ &
= \sigma(\ebar) = \sT S_0(\ebar)\ .
\end{flalign}
The first moment is
\begin{flalign}
\langle \e_s^1\sigma \rangle & = \int d\Omega_s \int d\e_s\ \e_s\ \frac{d\sigma}{d\e_sd\Omega_s}
\nonumber \\ &
= \g\sT\Biggr\{ S_0(\ebar) 
- \frac{3}{8\ebar^3}\Biggr[ \frac{\ebar^2}{3} \left( \frac{(1+2\ebar)^3-1}{(1+2\ebar)^3} \right)  
\nonumber \\ &
+ \frac{2\ebar(\ebar^2-\ebar-1)}{1+2\ebar} + \ln(1+2\ebar)
\Biggr] \Biggr\}\ 
\end{flalign}
\citep{dermer93,dermer09_book}.  Calculation of these and higher-order moments
were performed by \citet{coppi90}.  The mean scattered photon energy
is
\begin{flalign}
\langle \e_s \rangle = \frac{\langle \e_s^1\sigma \rangle }{\langle \epsilon_s^0\sigma \rangle}\ .
\end{flalign}
A particularly useful quantity is the mean scattered photon energy as a fraction of the incident
electron energy, 
\begin{flalign}
\label{S1}
\langle y\rangle & = \frac{\langle \e_s \rangle}{\g} 
= \frac{\langle \e_s^1\sigma \rangle }{\g\langle \epsilon_s^0\sigma \rangle}
\nonumber \\
& = S_1(\ebar)
\end{flalign}
where $S_1(x)$ is a function that can be computed numerically.  This
is a function only of the parameter $\ebar = \g\e(1-\cos\psi)$, a
useful property that I will exploit in Section \ref{general}.  The
function $S_1(x)$ and its asymptotes
\begin{equation}
\label{S1asympt}
S_1(x) \rightarrow \left\{ \begin{array}{ll}
x & x\ll 1 \\
1-4/[3(\ln(2x)+1/2)] & x \gg 1 
\end{array} \right.\ 
\end{equation}
are plotted in Figure \ref{S1plot}.  

\begin{figure}
\vspace{10.0mm} 
\epsscale{1.0} 
\plotone{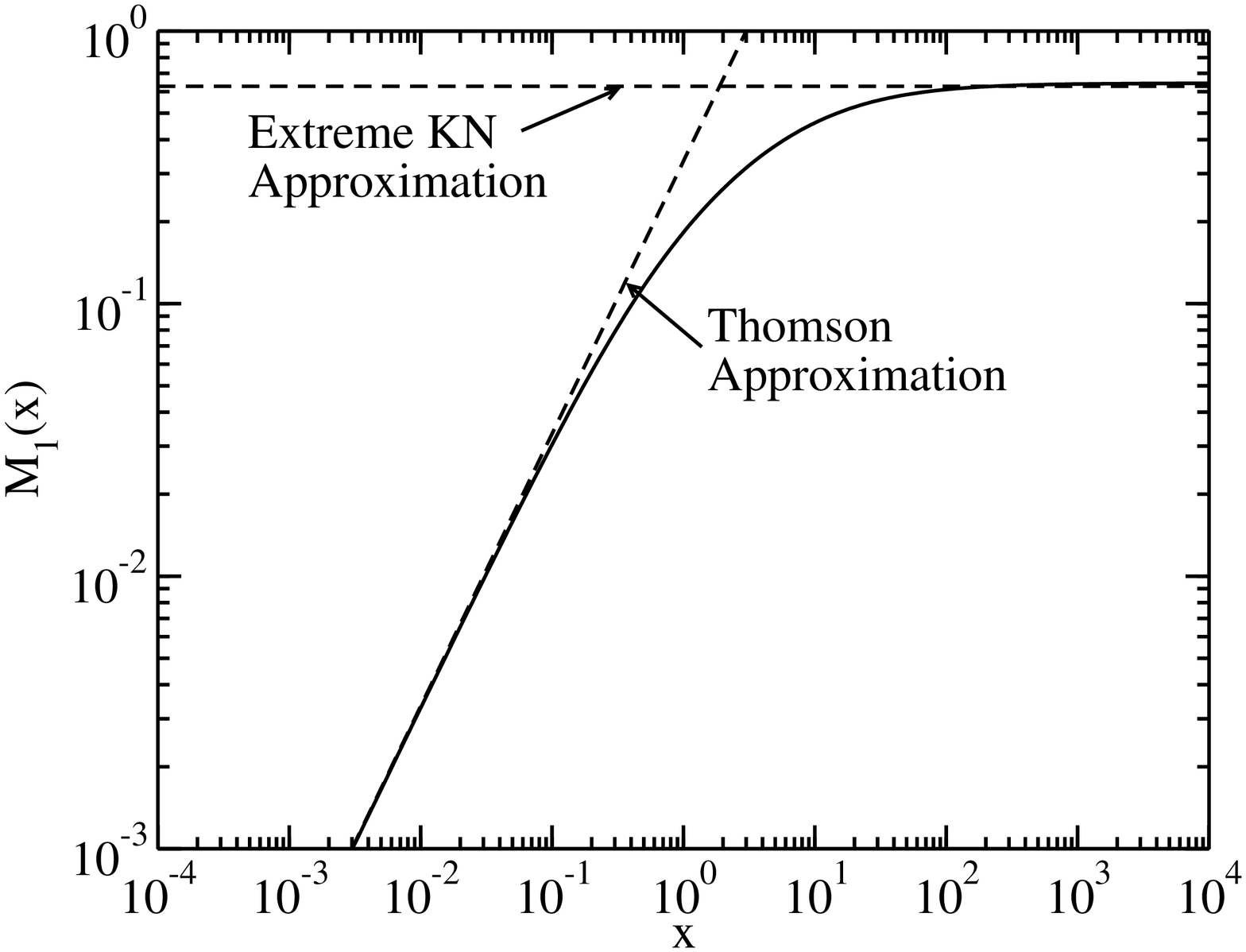}
\caption{The function $M_1(x)$ defined by Equation (\ref{M1}) and approximations from
Equation (\ref{M1asympt}) for $x\ll 1$ and $x\gg 1$.
}
\label{M1plot}
\vspace{2.2mm}
\end{figure}

It will be useful to find the angle-averaged mean photon energy as a fraction of 
incident electron energy.
This can be found from
\begin{flalign}
\label{M1_1}
\langle y \rangle_{\mu,\phi} & = \frac{\meanesangle}{\g} 
\nonumber \\
& = \frac{ \int^\infty_0 dy\ y\ \int^{2\pi}_0 d\phi \int^{1}_{-1} d\mu\ (1-\mu)\ \frac{d\sigma}{d\e_s}  }
{\int^\infty_0 dy\ \int^{2\pi}_0 d\phi \int^{1}_{-1} d\mu\ (1-\mu)\ \frac{d\sigma}{d\e_s}  }
\nonumber \\
& = M_1(4\g\epsilon) 
\ .
\end{flalign}
The integrals over $\mu$ and $\phi$ in Equation (\ref{M1_1}) have been
performed by \citet{jones68}, with corrections given by
\citet{blumen70}\footnote{An alternative but equivalent version of
the result of this integration is given by \citet{moderski05}.} so that
\begin{flalign}
\label{M1}
M_1(x) = \frac{ \int^{x/(1+x)}_0 dy\ y\ J_C(x,y) }{\int^{x/(1+x)}_0 dy\ J_C(x,y) }
\end{flalign}
where
\begin{flalign}
J_C(x,y) & = 2w\ln w + (1+2w)(1-w) 
\nonumber \\
& + \frac{1}{2}\frac{(xw)^2}{1+xw}(1-w)\ ,
\end{flalign}
in which
\begin{flalign}
w = \frac{y}{x(1-y)}\ .
\end{flalign} 
Analogous to $S_1(x)$, $M_1(x)$ is a function of only $x=4\g\e$, a
useful property exploited by \citet{moderski05} and by myself in
Section \ref{isotropic}.  It has the asymptotes
\begin{equation}
\label{M1asympt}
M_1(x) \rightarrow \left\{ \begin{array}{ll}
x/3 & x\ll 1 \\
0.6279 & x \gg 1 
\end{array} \right.\ .  
\end{equation}
The function $M_1(x)$ and its asymptotes are plotted in Figure
\ref{M1plot}.

\subsection{Delta Function Scattering Approximation for General Seed Radiation Field}
\label{general}

\citet{moderski05} used a $\delta$-function approximation for Compton
scattering valid for isotropic seed photon fields and for point source
seed photon fields.  In this section I generalize this for a
photon field with any angular distribution. 

\begin{figure}
\vspace{10.0mm} 
\epsscale{1.0} 
\plotone{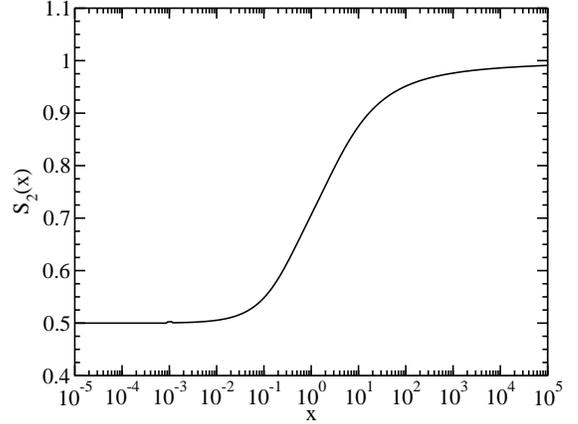}
\caption{The function $S_2(x)$ given by Equation (\ref{S2}).
}
\label{S2plot}
\vspace{2.2mm}
\end{figure}

A $\delta$-function approximation, valid for any seed photon field, is
\begin{flalign}
\label{deltaapprox1}
\frac{d\sigma}{d\e_s} & \approx \sigma(\ebar)
\delta\left(\e_s-\frac{5}{4}\langle\e_s\rangle\right)
\nonumber \\
& \approx \sigma(\ebar)\delta\left(\e_s-\frac{5}{4}\g S_1(\ebar)\right) \ .
\end{flalign}
The factor 5/4 comes from numerical experimentation, it leads to the
best reproduction of the full calculation.  The $\delta$-function
above can be rearranged so that
\begin{flalign}
\label{deltaapprox2}
\frac{d\sigma}{d\e_s} \approx 
\frac{4\langle\g\rangle}{5\e_s} \sigma(\ebar)S_2(\ebar) \delta\left(\g-\langle\g\rangle\right)
\end{flalign}

where $\langle\g\rangle$ is obtained by solving
\begin{flalign}
\label{esg}
\frac{\e_s}{\langle\g\rangle} = \frac{5}{4}S_1\left(\langle\g\rangle\e(1-\cos\psi)\right)\ .
\end{flalign}
In general, Equation (\ref{esg}) must be solved numerically, for example using the 
routine \verb ZBRENT \ from \citet{press92}, although
for the asymptotes given by Equation (\ref{S1asympt}) it has the
closed-form solutions
\begin{equation}
\langle \g \rangle \rightarrow \left\{ \begin{array}{ll}
\sqrt{ \frac{4\e_s}{5\e(1-\cos\psi)} } & x\ll 1 \\
\frac{4}{5}\e_s  & x \gg 1 
\end{array} \right. \ .
\end{equation}
Hereafter, in all calculations where the $\delta$-approximation is
used, I simplify the notation and allow
$\langle\g\rangle\rightarrow\g$.  In Equation (\ref{deltaapprox2}),
\begin{flalign}
\label{S2}
S_2(x) = \frac{d\ln(x)}{d\ln(xS_1(x))}\ ,
\end{flalign}
which has the asymptotes
\begin{equation}
\label{S2asympt}
S_2(x) \rightarrow \left\{ \begin{array}{ll}
1/2 & x \ll 1 \\
1   & x \gg 1 
\end{array} \right.\ 
\end{equation}
and is plotted in Figure \ref{S2plot}.
Putting the approximation Equation (\ref{deltaapprox2}) in Equation
(\ref{emissivity2}) gives
\begin{flalign}
\label{emissivity3}
\e_s\, j(\e_s,\Omega_s) & \approx \frac{4}{5}\sT c \e_s 
\int_0^{2\pi} d\phi \int_{-1}^1 d\mu 
\nonumber \\ & \times
\int \frac{d\e}{\e^2} u(\e,\Omega) 
n_e(\g,\Omega_s) 
\nonumber \\ & \times
S_3(\g\e(1-\cos\psi)) 
\end{flalign}
for $\g_{\rm min}<\g$, and
\begin{flalign}
\label{emissivity4}
\e_s\, j(\e_s,\Omega_s) & \approx \frac{4}{5}\sT c \e_s 
\int_0^{2\pi} d\phi \int_{-1}^1 d\mu
\nonumber \\ & \times
\int \frac{d\e}{\e^2} u(\e,\Omega) n_e(\g_{\rm min},\Omega_s) 
\nonumber \\ & \times
S_3(\g_{\rm min}\e(1-\cos\psi))\left(\frac{\g}{\g_{\rm min}}\right)^2 \ 
\end{flalign}
for $\g<\g_{\rm min}$.  
Here
\begin{flalign}
S_3(x) = x S_0(x)S_2(x)\ ,
\end{flalign}
\begin{flalign}
u(\e,\Omega) = m_ec^2\e n_{ph}(\e,\Omega)\ ,
\end{flalign}
and
\begin{flalign}
\g_{\rm min} = \frac{\e_s}{2}\left[ 1 + \sqrt{ 1 + \frac{2}{\e\e_s(1-\cos\psi)}}\right]\ .
\end{flalign}

\subsection{Delta Function Scattering Approximation for Isotropic Seed Radiation Field}
\label{isotropic}

In this section I derive the result for isotropic radiation fields,
reproducing the result from \citet{moderski05}.

For an isotropic seed photon field, 
\begin{flalign}
u(\e,\Omega) = \frac{u(\e)}{4\pi}\ ,
\end{flalign}
the spherical symmetry implies that the result is independent of
$\mu_s$.  Since one can then choose any value for $\mu_s$, one can 
choose $\mu_s=1$, so that
\begin{flalign}
\label{emissiv_iso2}
\e_s\, j(\e_s) & = \frac{c \e_s^2}{(4\pi)^2} 
\int^\infty_0 \frac{d\e}{\e}\ u(\e)
\int^\infty_1 d\g\ n_e(\g) 
\nonumber \\ & \times
\int^{2\pi}_0 
d\phi \int^{1}_{-1} d\mu\ 
(1-\mu)\ \frac{d\sigma}{d\e_s}\ .
\end{flalign}
Here one can use the approximation 
\begin{flalign}
\label{isodeltaapprox}
& \int^{2\pi}_0 d\phi \int^{1}_{-1} d\mu\ 
(1-\mu)\ \frac{d\sigma}{d\e_s} 
\nonumber \\ 
& \approx  \int^{2\pi}_0 d\phi \int^{1}_{-1} d\mu\ (1-\mu)\ 
\nonumber \\ & \times
\sigma_C(\g\e(1-\mu))
\delta\left(\e_s-\frac{3}{2} \meanesangle\right)
\nonumber \\ 
& \approx \frac{3\pi\sT}{4\g^2\e^2} M_0(4\g\e)
\delta\left(\e_s-\frac{3}{2} \g M_1(4\g\e)\right)
\end{flalign}
where the factor $3/2$ comes from numerical experimentation.

\begin{figure}
\vspace{10.0mm} 
\epsscale{1.0} 
\plotone{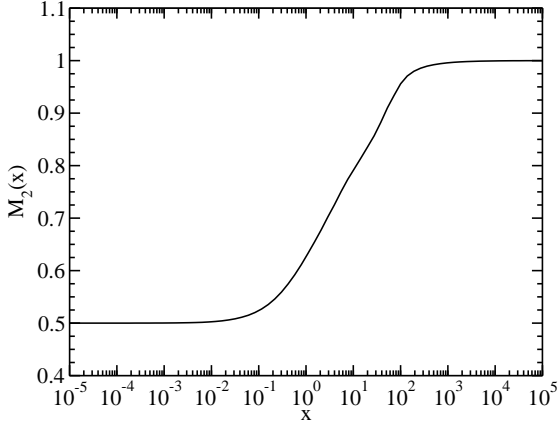}
\caption{The function $M_2(x)$ given by Equation (\ref{M2}).
}
\label{M2plot}
\vspace{2.2mm}
\end{figure}

Substituting Equation (\ref{isodeltaapprox}) into Equation
(\ref{emissiv_iso2}) and performing the integrals over $\mu$, $\phi$, and $\g$ results in
\begin{flalign}
\label{emissiv_iso4}
\e_s\, j(\e_s) = \frac{c\sT \e_s}{32\pi} \int^\infty_0 \frac{d\e}{\e^3}\ u(\e)
\frac{n_e(\g)}{\g}M_3(4\g\e)
\end{flalign}
for $\g_{\rm min}<\g$, and
\begin{flalign}
\label{emissiv_iso5}
\e_s\, j(\e_s) & = \frac{c\sT \e_s}{32\pi} \int^\infty_0 \frac{d\e}{\e^3}\ u(\e)
\frac{n_e(\g_{\rm min})}{\g_{\rm min}}
\nonumber \\ & \times
M_3(4\g_{\rm min}\e)\left(\frac{\g}{\g_{\rm min}}\right)^{3/2}
\end{flalign}
for $\g < \g_{\rm min}$, where 
\begin{flalign}
M_3(x) = M_0(x)M_2(x)\ ,
\end{flalign}
\begin{flalign}
\label{M2}
M_2(x) = \frac{d\ln(x)}{d\ln(xM_1(x))}\ ,
\end{flalign}
and
\begin{flalign}
\g_{\rm min} = \frac{\e_s}{2}\left[ 1 + \sqrt{1 + \frac{1}{\e\e_s} } \right]\ .
\end{flalign}
The function $M_2(x)$ has the asymptotes
\begin{equation}
M_2(x) \rightarrow \left\{ \begin{array}{ll}
1/2 & x \ll 1 \\
1   & x \gg 1 
\end{array} \right.\ .
\end{equation}
and is plotted in Figure \ref{M2plot}.
In general, one must solve
\begin{flalign}
\label{gammaiso}
\frac{\e_s}{\g} = \frac{3}{2}M_1(4\g\e)
\end{flalign}
for $\g$ numerically.  For the asymptotes given by Equation (\ref{M1asympt}), 
Equation (\ref{gammaiso}) can be solved analytically, so that
\begin{equation}
\g \rightarrow \left\{ \begin{array}{ll}
\sqrt{ \e_s/(2\e)} & x\ll 1 \\
\e_s/(0.9419)  & x \gg 1 
\end{array} \right. \ .
\end{equation}

\section{External Compton Scattering}
\label{ECsection}

Consider a homogeneous plasma ``blob'' moving at angle
$\theta_{s}=\arccos\mu_{s}$ to our line of sight with relativistic
speed $\beta c$ relative to a stationary black hole and galaxy giving
it a bulk Lorentz factor $\G = (1-\beta^2)^{-1/2}$ and Doppler factor
$\dD = [\G(1-\beta\mu_{s})]^{-1}$.  This represents a compact feature
in a relativistic jet of a blazar.  The blazar has cosmological
redshift $z$ giving it a luminosity distance $d_L(z)$.  Quantities in
the frame comoving with the blob are denoted by a prime, quantities in
the ``stationary'' frame of the black hole and host galaxy are
unmarked, and quantities in the observer frame are marked with an
``obs''.  The observed $\nu F_{\nu}$ flux is
\begin{flalign}
\label{fobsgen0}
f_{\e_s^{\rm obs}} = \frac{V_b \e_s\, j(\e_s,\Omega_s)}{d_L^2}
\end{flalign}
where $\e_s^{\rm obs} = \e_s/(1+z)$, and $V_b$ is the volume of the plasma
blob in the stationary frame.  The Compton-scattering of a stationary
radiation field external to the blob can be calculated, following
\citet{georgan01} and \citet{dermer09}, by transforming an isotropic
comoving electron distribution to the stationary frame, so that
\begin{flalign}
N_e(\g,\Omega_e) = \frac{\dD^3 N\p_e(\g/\dD)}{4\pi}
\end{flalign}
where $N_e(\g,\Omega_e) = V_b n_e(\g,\Omega_e)$.  Using this with
Equations (\ref{emissivity3}) and (\ref{emissivity4}) 
for an arbitrary seed photon energy density
one gets
\begin{flalign}
\label{fobsgen1}
f_{\e_s^{\rm obs}} & = \frac{c \sT \e_s \dD^3}{5\pi d_L^2}
\int_0^{2\pi} d\phi \int_{-1}^{1} d\mu
\nonumber \\ & \times
\int_0^\infty \frac{d\e}{\e^2} u(\e,\Omega) N\p_e(\g/\dD)
\nonumber \\ & \times
S_3(\g\e(1-\cos\psi))\ 
\end{flalign}
for $\g_{\rm min}<\g$, and
\begin{flalign}
\label{fobsgen2}
f_{\e_s^{\rm obs}} & = \frac{c \sT \e_s \dD^3}{5\pi d_L^2}
\int_0^{2\pi} d\phi \int_{-1}^{1} d\mu
\nonumber \\ & \times
\int_0^\infty \frac{d\e}{\e^2} u(\e,\Omega) N\p_e(\g_{\rm min}/\dD)
\nonumber \\ & \times 
S_3(\g_{\rm min}\e(1-\cos\psi))
\left(\frac{\g}{\g_{\rm min}}\right)^2 \ 
\end{flalign}
for $\g<\g_{\rm min}$.  

For an isotropic seed photon energy density, using
Equations (\ref{emissiv_iso4}) and (\ref{emissiv_iso5}) one gets
\begin{flalign}
\label{fluxiso1}
f_{\e_s^{\rm obs}} & = \frac{c\sT \dD^3\e_s^{\rm obs}(1+z)}{32\pi d_L^2}
\int^\infty_0 \frac{d\e}{\e^3}u(\e) 
\nonumber \\ & \times
\frac{N^\prime_e(\g/\dD)}{\g}M_3(4\g\e)
\end{flalign}
for $\g_{\rm min}<\g$, and
\begin{flalign}
\label{fluxiso2}
f_{\e_s^{\rm obs}} & = \frac{c\sT \dD^3\e_s^{\rm obs}(1+z)}{32\pi d_L^2}
\int^\infty_0 \frac{d\e}{\e^3}u(\e) 
\nonumber \\ & \times
\frac{N^\prime_e(\g_{\rm min}/\dD)}{\g_{\rm min}}M_3(4\g_{\rm min}\e)
\left(\frac{\g}{\g_{\rm min}}\right)^{3/2}
\end{flalign}
for $\g<\g_{\rm min}$.

\subsection{Isotropic Monochromatic Radiation Field}
\label{isomonosection}

External radiation fields such as the CMB, or those from the BLR, or dust torus, 
are often represented by an isotropic, monochromatic external radiation
field.  In this case, 
\begin{flalign}
\label{u_isomono}
u(\e) = u_0\delta(\e-\e_0)\ .
\end{flalign}
Using Equations (\ref{fluxiso1}) and (\ref{fluxiso2}) one gets
\begin{flalign}
\label{fluxisomono1}
f_{\e_s^{\rm obs}} & = \frac{c\sT \dD^3\e_s^{\rm obs}(1+z)}{32\pi d_L^2}
 \frac{u_0}{\e_0^3} 
\nonumber \\ & \times
\frac{N^\prime_e(\g/\dD)}{\g}M_3(4\g\e_0)
\end{flalign}
for $\g_{\rm min}<\g$, and
\begin{flalign}
\label{fluxisomono2}
f_{\e_s^{\rm obs}} & = \frac{c\sT \dD^3\e_s^{\rm obs}(1+z)}{32\pi d_L^2}
\frac{u_0}{\e_0^3} 
\nonumber \\ & \times
\frac{N^\prime_e(\g_{\rm min}/\dD)}{\g_{\rm min}}M_3(4\g_{\rm min}\e_0)
\left(\frac{\g}{\g_{\rm min}}\right)^{3/2}
\end{flalign}
for $\g<\g_{\rm min}$.
In the Thomson regime, $4\g\e_0\ll 1$, Equation (\ref{fluxisomono1}) reduces to
\begin{flalign}
f_{\e_s^{\rm obs}} \rightarrow \frac{\dD^4}{6\pi d_L^2} c \sT \g^{\p 3}_T N\p_e(\gp_T)\ ,
\nonumber \\ 
\gp_T = \frac{1}{\dD} \sqrt{ \frac{\e_s^{\rm obs}(1+z)}{2\e_0} } \ ,
\end{flalign}
which is the approximation given in Equation (6.109) of
\citet{dermer09_book}.

\subsection{Monochromatic Point Source Radially Behind Jet}
\label{ptsourcesection}

Scattering of photons from an accretion disk is sometimes approximated
by the scattering of a monochromatic point source radially behind the
jet; that is, at $\mu=1$ \citep[e.g.,][]{dermer92}.  In this case
the external energy density is
\begin{flalign}
\label{u_ptsrc}
u(\e,\Omega) = \frac{L_0}{4\pi r^2 c}\frac{\delta(\mu-1)}{2\pi} \delta(\e-\e_0)\ .
\end{flalign}
Then Equations (\ref{fobsgen1}) and (\ref{fobsgen2}) give
\begin{flalign}
\label{fluxbehindjet}
f_{\e_s^{\rm obs}} & = \frac{\e_s L_{0}\sT\dD^3}{16\pi^2 r^2 d_L^2\e_0^2}
S_3[\g\e(1-\mu_s)]
\nonumber \\ & \times
N\p_e(\g/\dD)
\end{flalign}
for $\g_{\rm min}<\g$, and
\begin{flalign}
\label{fluxbehindjet2}
f_{\e_s^{\rm obs}} & = \frac{\e_s L_{0}\sT\dD^3}{16\pi^2 r^2 d_L^2\e_0^2}
S_3[\g_{\rm min}\e(1-\mu_s)]
\nonumber \\ & \times
N\p_e(\g_{\rm min}/\dD)\left(\frac{\g}{\g_{\rm min}}\right)^2
\end{flalign}
for $\g<\g_{\rm min}$.

\subsection{Shakura-Sunyaev Accretion Disk}
\label{disksection}

The energy density of photons from an accretion disk from the solution of \citet{shakura73}
can be approximated as
\begin{flalign}
\label{u_disk}
u(\e) = \frac{3}{16\pi^2 c}\frac{\ell_{\rm Edd}L_{\rm Edd}R_g}{\eta R^3}\varphi(R)
\delta(\e-\e_0(R))
\end{flalign}
\citep{dermer02,dermer09} where
\begin{flalign}
\label{e0diskeqn}
\e_0(R) = 2.7\times10^{-4}\left(\frac{\ell_{\rm Edd}}{M_8\eta} \right)^{1/4} \left(\frac{R}{R_g}\right)^{-3/4}\ ,
\end{flalign}
\begin{flalign}
R_g = \frac{GM}{c^2} \approx 1.5\times10^{13}M_8\ \cm\ ,
\end{flalign}
\begin{flalign}
\varphi(R) = \sqrt{ 1 - \frac{R_{\rm in}}{R} }\ ,
\end{flalign}
$\ell_{\rm Edd}=L_{\rm disk}/L_{\rm Edd}$, $L_{\rm disk}$ is the disk luminosity,
$L_{\rm Edd} = 1.26\times10^{46}M_8$ is the Eddington luminosity, $\eta$
is the accretion efficiency, $M_{\rm BH}=10^8 M_8 M_{\odot}$ is the
black hole mass, and $R_{\rm in}$ is the inner radius of the accretion
disk.  For a nonrotating Schwarzschild black hole, one expects that
$R_{\rm in}=6R_g$.  For a rotating Kerr black hole, the inner radius is
given by
\begin{flalign}
R_{\rm in} & = R_g \Biggr\{ 3 + A_2 - 
\nonumber \\ &
\frac{|a|}{a}\left[(3-A_1)(3+A_1+2A_2)\right]^{1/2} \Biggr\}
\end{flalign}
\citep[e.g.,][]{zhang97} where
\begin{flalign}
A_1 & = 1 + (1-a^2)^{1/3}
\nonumber \\ & \times
[(1+a)^{1/3} + (1-a)^{1/3}]\ ,
\end{flalign}
\begin{flalign}
A_2 = \sqrt{3a^2+A_1^2} \ , 
\end{flalign}
and $a$ ($-1 \le a \le 1$) is the black hole spin.

The outer disk radius can be estimated as where the disk's
self-gravity dominates over the black hole gravity
\citep{laor89,netzer15}.  In this paper I simply use a constant
value for the outer disk radius of $R_{\rm out} = 200R_g$.

The Compton-scattered flux with the accretion disk as the seed photon
source is, for $\g_{\rm min}<\g$,
\begin{flalign}
\label{fluxECeqn}
f_{\e_s^{\rm obs}} & = \frac{3\e_s \dD^3\ell_{\rm Edd}L_{\rm Edd} R_g \sT}{64\pi^3 d_L^2\eta r^3}
\int_0^{2\pi}d\phi\ 
\nonumber \\ & \times
\int_{\mu_{\rm min}}^{\mu_{\rm max}}d\mu \ \frac{\varphi(\mu)}{(\mu^{-2}-1)^{3/2}\e_0(\mu)^2}
\nonumber \\ & \times
S_3[\g\e_0(\mu)(1-\cos\psi)]  N\p_e(\g/\dD)
\end{flalign}
and, for $\g<\g_{\rm min}$, 
\begin{flalign}
\label{fluxECeqn2}
f_{\e_s^{\rm obs}} & = \frac{3\e_s \dD^3\ell_{\rm Edd}L_{\rm Edd} R_g \sT}{64\pi^3 d_L^2\eta r^3}
\int_0^{2\pi}d\phi\ 
\nonumber \\ & \times
\int_{\mu_{\rm min}}^{\mu_{\rm max}}d\mu \ 
\frac{\varphi(\mu)}{(\mu^{-2}-1)^{3/2}\e_0(\mu)^2}\ ,
\nonumber \\ & \times
S_3[\g_{\rm min}\e_0(\mu)(1-\cos\psi)]  N\p_e(\g_{\rm min}/\dD)
\nonumber \\ & \times
\left(\frac{\g}{\g_{\rm min}}\right)^2\ ,
\end{flalign}
where
\begin{flalign}
\mu_{\rm min} = \frac{1}{\sqrt{ 1 + (R_{\rm out}/r)^2}}
\end{flalign}
and
\begin{flalign}
\mu_{\rm max} = \frac{1}{\sqrt{ 1 + (R_{\rm in}/r)^2}}\ .
\end{flalign}

\begin{figure}
\vspace{2.2mm} 
\epsscale{1.0} 
\plotone{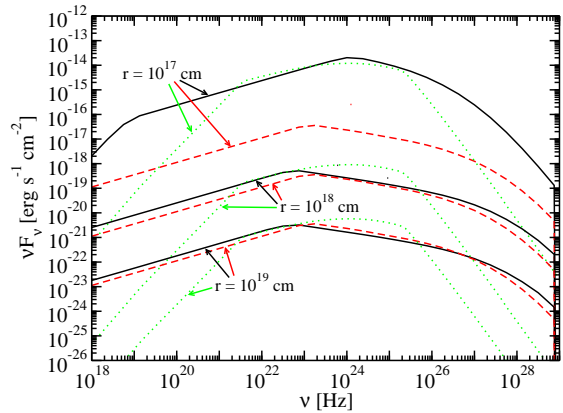}
\caption{The SED of the Compton-scattered
Shakura-Sunyaev accretion disk radiation (Equation [\ref{fluxECeqn}]
and [\ref{fluxECeqn2}]; solid black curves), the monochromatic point
source behind the jet (Equation [\ref{fluxisomono1}] and
[\ref{fluxisomono2}]; dashed red curves); and the external isotropic
monochromatic radiation field (Equation [\ref{fluxbehindjet}] and
[\ref{fluxbehindjet2}]; dotted green curves) for various distances
from the black hole $r$.  The scattering of the external isotropic
radiation field calculation had its energy density modified until its
normalization approximately matched the scattered Shakura-Sunyaev
calculation. }
\label{fluxECdisk}
\vspace{2.2mm}
\end{figure}

In Figure \ref{fluxECdisk} I plot the Compton-scattered
Shakura-Sunyaev accretion disk field, along with the Compton-scattered
external isotropic monochromatic radiation field and the monochromatic
point source from behind.  The electron distribution used is a broken
power-law, given by
\begin{equation}
N\p_e(\gp) = N_e(\gp_b) \left\{ \begin{array}{ll}
(\gp/\gp_b)^{-p_1} & \gp \le \gp_b \\
(\gp/\gp_b)^{-p_2} & \gp_b < \gp
\end{array}
\right. \ .
\end{equation}
The parameters for the ``baseline'' model for the comparison are given
in Table \ref{compareparamtable}; the parameter $r$ is varied from the
baseline model.  In the monochromatic approximations, the seed photon
energies are $\e_0(R=8.17R_g)$, as calculated from Equation
(\ref{e0diskeqn}).  For the external isotropic radiation field
calculation, the energy density of the scattered field ($u_0$) has
been adjusted so that the normalization is approximately consistent
with that of the scattered Shakura-Sunyaev disk field calculation.  It
is apparent that the isotropic monochromatic field is a poor
approximation for the Shakura-Sunyaev disk field for scattering
calculations.  The monochromatic point source from behind is a
reasonable approximation for the Shakura-Sunyaev disk field at large
$r$.

\begin{deluxetable}{lcc}
\tabletypesize{\scriptsize}
\tablecaption{Parameters for Numerical Tests}
\tablewidth{0pt}
\tablehead{
\colhead{Parameter} &
\colhead{Symbol} &
\colhead{Value}
}
\startdata
Lorentz factor               &    $\G$                & $40$ \\
Doppler factor               &    $\dD$               & $40$ \\
Magnetic Field [G]           &    $B$                 & $0.56$  \\
\hline
ED\tablenotemark{a}\ minimum Lorentz factor    &    $\gp_{1}$      & $20$ \\
ED\tablenotemark{a}\ maximum Lorentz factor    &    $\gp_{2}$      & $5\times10^7$ \\
ED\tablenotemark{a}\ peak Lorentz factor    &    $\gp_{b}$      & $10^4$ \\
ED\tablenotemark{a}\ first index        &    $p_1$                 & $2$  \\
ED\tablenotemark{a}\ second index        &    $p_2$                 & $3.5$  \\
\hline
Black hole mass [$M_{\odot}$]      &    $M_{\rm BH}$    & $1.2\times10^9$ \\
Gravitational radius [cm]          &    R$_g$           & $1.8\times10^{14}$ \\
Disk luminosity [$\erg\ \s^{-1}$]  &    $L_{\rm disk}$  & $2\times10^{46}$  \\
Accretion efficiency               &    $\eta$          & $1/12$ \\
Inner disk radius [$R_g$]          &    $R_{\rm in}$    & $6$        \\
Outer disk radius [$R_g$]          &    $R_{\rm out}$   & $200$  \\
BL\tablenotemark{b} dimensionless energy            &    $\e_{\rm li}$   & $2\times10^{-5}$ \\
BL\tablenotemark{b} scattering fraction             &    $\xi_{\rm li}$  & 0.024       \\
BL\tablenotemark{b} radius [cm]                     &    $R_{\rm li}$    & $10^{17}$       \\
DT Temperature                     & $\e_{\rm dt}$  & $10^3$ \\
DT\tablenotemark{b} scattering fraction             & $\xi_{\rm dt}$ & $0.1$ \\
DT\tablenotemark{b} inner radius [cm]               & $R_{\rm dt,1}$ & $1.6\times10^{19}$ \\
DT\tablenotemark{b} outer radius [cm]               & $R_{\rm dt,2}$ & $1.6\times10^{20}$ \\
DT\tablenotemark{b} gradient                        & $\zeta$        & $1.0$ \\
blob distance from black hole [cm] &    $r$             & $10^{17}$
\enddata
\tablenotetext{a}{Electron Distribution}
\tablenotetext{b}{Broad Line}
\tablenotetext{c}{Dust Torus}
\label{compareparamtable}
\end{deluxetable}

\subsection{Reprocessing of Accretion Disk Radiation}

\begin{figure}
\vspace{2.2mm} 
\epsscale{1.0} 
\plotone{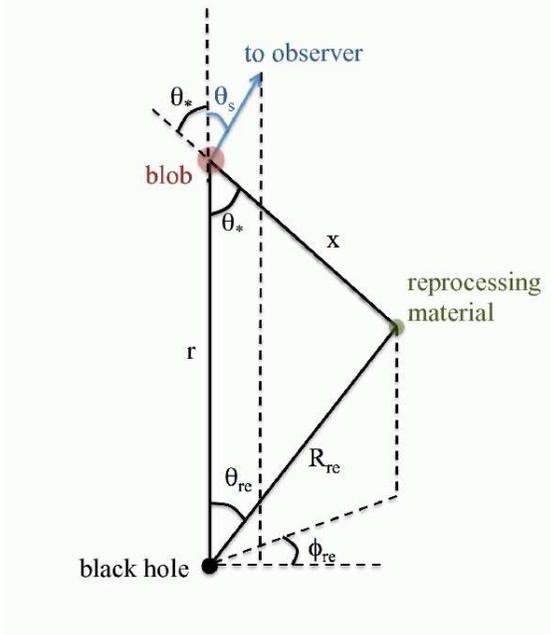}
\caption{An illustration of the geometry of reprocessed emission as a
seed photon source for Compton scattering by electrons in the blob.}
\label{geometry}
\vspace{2.2mm}
\end{figure}

The accretion disk radiation described above can be absorbed and
re-radiated as line emission in the broad-line region (BLR), or
absorbed and re-radiated in the infrared as approximately a blackbody
by a dust torus.  The geometry for this setup is illustrated in Figure
\ref{geometry}.  Assuming that the disk radiation originates at the
origin (the location of the central black hole), and that the
reprocessed emission is emitted isotropically so that it has
emissivity $j_{\rm re}(\e,\Omega_{\rm re}; R_{\rm re})$, then the
energy density of the reprocessed emission is
\begin{flalign}
u_{\rm re}(\e) = \int dV_{\rm re} \frac{j_{\rm re}(\e,\Omega_{\rm re}; R_{\rm re})}{4\pi x^2c}\ 
\end{flalign}
where $R_{\rm re}$ is the distance between the central
source and the reprocessing medium, and $x$ is the distance between
the reprocessing medium and the emitting blob.  Recall that $r$ is the
distance between central source and the emitting blob, so that
\begin{flalign}
x^2 = R_{\rm re}^2 + r^2 - 2rR_{\rm re}\mu_{\rm re}\ .
\end{flalign}
The energy density per solid angle [where $u(\e)=\int d\Omega\
u(\e,\Omega)$] can be found by imposing $\delta$-function constraints
\citep{boett95,dermer09},
\begin{flalign}
\label{ure}
u_{\rm re}(\e,\Omega) & = \int dV_{\rm re} \frac{j_{\rm re}(\e,\Omega_{\rm re}; R_{\rm re})}{4\pi x^2c}\ 
\nonumber \\ &\times
\delta(\phi-\phi_{\rm re}) \delta(\mu-\mu_*)
\nonumber \\ 
& = \frac{1}{4\pi c}\int^{2\pi}_0 d\phi_{\rm re} \int^{1}_{-1}d\mu_{\rm re} \int^{\infty}_{0} dR_{\rm re}\ 
\nonumber \\ &\times
\left(\frac{R_{\rm re}}{x}\right)^2\ j_{\rm re}(\e,\Omega_{\rm re}; R_{\rm re})\ 
\nonumber \\ &\times
\delta(\phi-\phi_{\rm re}) \delta(\mu-\mu_*)
\end{flalign}
where
\begin{flalign}
\label{mustar}
\mu_*^2 = 1 - \left(\frac{R_{\rm re}}{x}\right)^2(1-\mu_{\rm re}^2)  \ .  
\end{flalign}

\subsection{Broad Line Region}
\label{BLRsection}

The BLR around black holes reprocesses disk radiation into line
emission that is Doppler-broadened by the BLR clouds' orbits around the
black hole.  Reverberation mapping indicates that individual lines
emit primarily in a relatively narrow distance from the black hole,
and that different lines are emitted at different radii
\citep[e.g.,][]{peterson99,kollat03,peterson14}.  Thus it seems
reasonable to assume, at least for the purposes of Compton scattering,
that each line is monochromatic, emitting photons with dimensionless
energy $\e_{\rm li}$ (i.e., ignoring the Doppler-broadening of the line)
and emitted at one distance from the BLR with radius $R_{\rm li}$.  I
consider two possible geometries for the BLR, a spherical shell and a
flattened annulus.  A simple empirical model for estimating the
luminosities (equivalently, $\xi_{\rm li}$) and radii ($R_{\rm li}$) of broad
lines is presented in Appendix \ref{BLRmodel}.

\subsubsection{Spherical Shell Broad Line Region}
\label{shellBLRsection}

I consider here that each line is emitted from an infinitesimally thin
spherical shell with radius $R_{\rm li}$ from the origin and
monochromatic dimensionless energy $\e_{\rm li}$.  In this case the
emissivity of the radiation from a single line is
\begin{flalign}
j_{\rm re}(\e, \Omega_{\rm re}; R_{\rm li}) = \frac{\xi_{\rm li} L_{\rm disk}}{4\pi R_{\rm li}^2}\delta(\e-\e_{\rm li})
\delta(R_{\rm re}-R_{\rm li})
\end{flalign}
where $\xi_{\rm li}$ is the fraction of disk radiation reprocessed by
the line.  Using this in Equation (\ref{ure}) gives
\begin{flalign}
\label{u_BLR}
u_{\rm re}(\e,\Omega) = \frac{\xi_{\rm li}L_{\rm disk}}{(4\pi)^2 c} \delta(\e-\e_{\rm li})
\int_{-1}^1 \frac{d\mu_{\rm re}}{x^2}
\ \delta(\mu-\mu_*)
\end{flalign}
for the line energy density.
The observed flux from Compton scattering BLR photons is then
\begin{flalign}
\label{fluxBLRshelleqn1}
f_{\e_s^{\rm obs}} & = \frac{\xi_{\rm li}L_{\rm disk}\sT\dD^3}{80\pi^3 d_L^2}\left(\frac{\e_s}{\e_{\rm li}^2}\right)
\int_0^{2\pi}d\phi
\nonumber \\ &\times
\int_{-1}^{1}\frac{d\mu_{\rm re}}{x^2} 
S_3[\g\e_0(1-\cos\bar{\psi})]N\p_e(\g/\dD)
\end{flalign}
for $\g_{\rm min}<\g$ and
\begin{flalign}
\label{fluxBLRshelleqn2}
f_{\e_s^{\rm obs}} & = \frac{\xi_{\rm li}L_{\rm disk}\sT\dD^3}{80\pi^3 d_L^2}\left(\frac{\e_s}{\e_{\rm li}^2}\right)
\int_0^{2\pi}d\phi
\nonumber \\ & \times
\int_{-1}^{1}\frac{d\mu_{\rm re}}{x^2} 
S_3[\g_{\rm min}\e_0(1-\cos\bar{\psi})]
\nonumber \\ &\times
N\p_e(\g_{\rm min}/\dD) \left(\frac{\g}{\g_{\rm min}}\right)^2
\end{flalign}
for $\g<\g_{\rm min}$ where
\begin{flalign}
\cos\bar{\psi} = \mu_*\mu_s + \sqrt{1-\mu_*^2}\sqrt{1-\mu_s^2}\cos\phi\ .
\end{flalign}

\subsubsection{Flattened Broad Line Region}
\label{ringBLRsection}

There is some evidence that the BLR is not a spherical shell, but
flatted \citep[e.g.,][]{mclure01,decarli11,shaw12}.  Thus I also
consider a model for the BLR where each line emits in an
infinitesimally thin ring with radius $R_{\rm li}$ oriented orthogonal
to the jet axis.  This can be computed by considering an additional
$\delta$-function constraint on $\mu_{\rm re}$, so that the emissivity
of a line is
\begin{flalign}
\label{BLRringemissiv}
j_{\rm re}(\e, \Omega_{\rm re}; R_{\rm li}) & = \frac{\xi_{\rm li}L_{\rm disk}}{4\pi R_{\rm li}^2}
\delta(R_{\rm re}-R_{\rm li})
\nonumber \\ & \times
\delta(\e-\e_{\rm li})\ \delta(\mu_{\rm re}-0)\ .
\end{flalign}
Using Equation (\ref{BLRringemissiv}) in Equation (\ref{ure}) gives
\begin{flalign}
\label{uBLRring}
u_{\rm re}(\e,\Omega) = \frac{ \xi_{\rm li}L_{\rm disk} }{(4\pi)^2 c x^2}
\delta(\mu-r/x) \delta(\e-\e_{\rm li})\ 
\end{flalign}
for the energy density, where
\begin{flalign}
x^2 = R_{\rm li}^2 + r^2\ .
\end{flalign}
The observed flux from Compton-scattering this radiation field is
\begin{flalign}
\label{fluxBLRringeqn1}
f_{\e_s^{\rm obs}} & = \frac{\e_s \xi_{\rm li}L_{\rm disk}\sT\dD^3}{80\pi^3\ x^2 d_L^2\e_{\rm li}^2}
\int^{2\pi}_0 d\phi\ 
\nonumber \\ & \times
S_3[\g\e(1-\cos\bar{\psi})]
N\p_e(\g/\dD)
\end{flalign}
for $\g_{\rm min}<\g$, and
\begin{flalign}
\label{fluxBLRringeqn2}
f_{\e_s^{\rm obs}} & = \frac{\e_s \xi_{\rm li}L_{\rm disk}\sT\dD^3}{80\pi^3\ x^2 d_L^2\e_{\rm li}^2}
\nonumber \\ & \times
\int^{2\pi}_0 d\phi\ S_3[\g_{\rm min}\e(1-\cos\bar{\psi})]
\nonumber \\ & \times
N\p_e(\g_{\rm min}/\dD)
\left(\frac{\g}{\g_{\rm min}}\right)^2
\end{flalign}
for $\g<\g_{\min}$, where
\begin{flalign}
\cos\bar{\psi} = \frac{r\mu_s}{x} + \sqrt{1-\frac{r^2}{x^2}}\sqrt{1-\mu_s^2}\cos\phi\ .
\end{flalign}

\begin{figure}
\vspace{2.2mm} 
\epsscale{1.0} 
\plotone{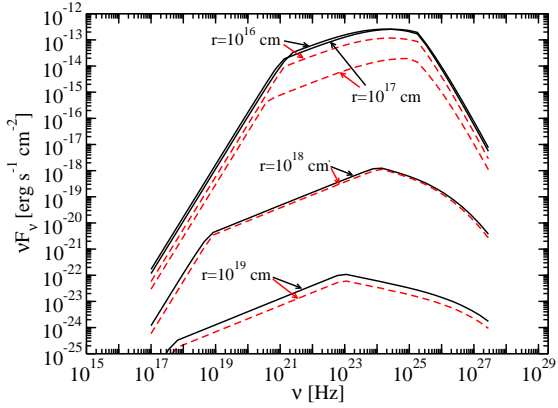}
\caption{The SED of the Compton-scattered BLR radiation
for a shell geometry (Equation [\ref{fluxBLRshelleqn1}] and
[\ref{fluxBLRshelleqn2}]; solid black curves), and for a flattened
geometry (Equation [\ref{fluxBLRringeqn1}] and
[\ref{fluxBLRringeqn2}]; dashed red curves) for various distances $r$
from the black hole. }
\label{fluxECBLR}
\vspace{2.2mm}
\end{figure}

In Figure \ref{fluxECBLR} I plot the Compton-scattered
BLR radiation field for both the shell and flattened BLR geometry.  I
use the same parameters as in Table \ref{compareparamtable}, while
varying the distance of the emitting region from the black hole ($r$).
In general, the flattened geometry gives a slightly lower flux
compared to the shell geometry.  This is most apparent at the distance
of the BLR line radius, $r=R_{\rm li}=10^{17}\ \cm$.  

\subsection{Dust Torus}
\label{dustsection}


Dust tori found in AGN emit approximately as blackbodies and are
thought to be oriented orthogonal to the jet axes, although its exact
geometry is uncertain.  Here I will use two approximations for the
dust torus, useful for Compton-scattering.  In both cases I will
approximate the dust torus blackbody as monochromatic with energy at
the peak of the blackbody distribution, $\e=2.7\Theta$, where
$\Theta=kT_{\rm dt}/(m_e c^2)$ is the dimensionless temperature of the
dust torus.  To avoid sublimation, the dust torus temperature $T_{\rm
dt}\la 2000\ \Kelvin$.  One could modify the expressions below to use
the full blackbody expression by allowing
$$
\delta(\e-2.7\Theta) \rightarrow \frac{15}{\pi^4\Theta^4} 
\frac{\e^3}{e^{\e/\Theta}-1}\ .
$$

\subsubsection{Ring Dust Torus}

Here I approximate the dust torus as an infinitesimally thin annulus
with radius $R_{\rm dt}$.  This is essentially the same approximation
as the ring BLR (Section \ref{ringBLRsection}).  For this case the
emissivity is
\begin{flalign}
j_{\rm re}(\e, \Omega; R_{\rm dt}) & = \frac{\xi_{\rm dt}L_{\rm disk}}{4\pi R_{\rm re}^2}
\delta(R_{\rm re}-R_{\rm dt})
\nonumber \\ & \times
\delta(\e-2.7\Theta)\ \delta(\mu_{\rm re}-0)\ .
\end{flalign}
Using this in Equation (\ref{ure}) gives
\begin{flalign}
\label{u_dust}
u_{\rm re}(\e,\Omega) = \frac{ \xi_{\rm dt}L_{\rm disk} }{(4\pi)^2 c x^2}
\delta(\mu-r/x) \delta(\e-2.7\Theta)\ 
\end{flalign}
for the dust torus energy density, where
\begin{flalign}
x^2 = R_{\rm dt}^2 + r^2\ .
\end{flalign}
The observed flux from Compton-scattering this radiation field is
\begin{flalign}
\label{fluxdustringeqn1}
f_{\e_s^{\rm obs}} = \frac{\e_s \xi_{\rm dt}L_{\rm disk}\sT\dD^3}{80\pi^3 (2.7\Theta)^2 x^2 d_L^2}
\int^{2\pi}_0 d\phi\ 
\nonumber \\ \times
S_3[2.7\g\Theta(1-\cos\bar{\psi})]
N\p_e(\g/\dD)
\end{flalign}
for $\g_{\rm min}<\g$, and
\begin{flalign}
\label{fluxdustringeqn2}
f_{\e_s^{\rm obs}} & = \frac{\e_s \xi_{\rm dt}L_{\rm disk}\sT\dD^3}{80\pi^3 (2.7\Theta)^2 x^2 d_L^2}
\int^{2\pi}_0 d\phi\ 
\nonumber \\ & \times
S_3[2.7\g_{\rm min}\Theta(1-\cos\bar{\psi})]
\nonumber \\ & \times
N\p_e(\g_{\rm min}/\dD)
\left(\frac{\g}{\g_{\rm min}}\right)^2
\end{flalign}
for $\g<\g_{\rm min}$, where
\begin{flalign}
\cos\bar{\psi} = \frac{r\mu_s}{x} + \sqrt{1-\frac{r^2}{x^2}}\sqrt{1-\mu_s^2}\cos\phi\ .
\end{flalign}
The dust torus radius can be estimated as the sublimation radius, 
\begin{flalign}
\label{Rdust}
R_{\rm dt} & = 3.5\times10^{18} \left(\frac{L_{\rm disk}}{10^{45}\ \erg\ \s^{-1}}\right)^{1/2}
\nonumber \\ & \times 
\left(\frac{T_{\rm dt}}{10^3\ \Kelvin}\right)^{-2.6}\ \cm\ 
\nonumber \\ 
& = 8\times10^5 \left(\frac{\ell_{\rm Edd}}{M_8}\right)^{1/2} T_{\rm dt,3}^{-2.6}\ R_g
\end{flalign}
\citep{nenkova08p2, sikora09} where $T_{\rm dt,3}=T_{\rm dt}/(10^3\
\Kelvin)$.

\subsubsection{Extended Dust Torus}

I have created another approximation to the dust torus based on the
dust modeling of \citet{nenkova08p1,nenkova08p2}.  I approximate the
dust torus as as a flattened disk with inner and outer radii, $R_{\rm
dt,1}$ and $R_{\rm dt,2}$, respectively.  The dust torus consists of a
number of dust clouds, each with cross sectional area $\Sigma_{\rm
cl}$.  The dust cloud number density scales as a power-law with
distance from the black hole ($R$), so that it is
\begin{flalign}
n_{\rm cl}(R) = n_{\rm cl,0}\left(\frac{R}{R_{\rm dt,1}}\right)^{-\zeta}\ 
H(R; R_{\rm dt,1}, R_{\rm dt,2})\ .
\end{flalign}
Other parameters of the dust torus ($\Sigma_{\rm cl}$ and $\Theta$) do
not vary with $R$ in my model, although in principle they could.  The fraction
of disk luminosity reprocessed by this dust torus is
\begin{flalign}
\xi_{\rm dt} = \int_0^\infty dR\ \Sigma_{\rm cl}\ n_{\rm cl}(R) = 
\Sigma_{\rm cl} n_{\rm cl,0} R_{\rm dt,eff}
\end{flalign}
where
\begin{equation}
R_{\rm dt,eff} = \left\{ \begin{array}{ll}
[R_{\rm dt,1}-R_{\rm dt,2}(R_{\rm dt,2}/R_{\rm dt,1})^{-\zeta}]/(\zeta-1) & \zeta \ne 1 \\
R_{\rm dt,1}\ln(R_{\rm dt,2}/R_{\rm dt,1}) & \zeta = 1
\end{array} \right.\ .
\end{equation}
With this formulation, the reprocessed emissivity is
\begin{flalign}
j_{\rm re}(\e, \Omega; R_{\rm re}) & = \frac{L_{\rm disk}}{4\pi R_{\rm re}^2}\Sigma_{\rm cl}n_{\rm cl}(R_{\rm re})
\nonumber \\ & \times
\delta(\e-2.7\Theta)\ \delta(\mu_{\rm re}-0)
\nonumber \\
& = \frac{\xi_{\rm dt}L_{\rm disk}}{4\pi R_{\rm re}^2 }
\frac{\delta(\e-2.7\Theta)\ \delta(\mu_{\rm re}-0)}{R_{\rm dt,eff}}\ ,
\end{flalign}
and Equation (\ref{ure}) gives
\begin{flalign}
\label{u_dust_ext}
u_{\rm re}(\e,\Omega) & = \frac{ \xi_{\rm dt} L_{\rm disk} \delta(\e-2.7\Theta)}{(4\pi)^2 c R_{\rm dt,eff}}
\int_{R_{\rm dt,1}}^{R_{\rm dt,2}} \frac{dR_{\rm re}}{x^2} 
\nonumber \\ & \times
\left(\frac{R_{\rm re}}{R_{\rm dt,1}}\right)^{-\zeta}
\delta(\mu-r/x) \ 
\end{flalign}
where
\begin{flalign}
x^2 = R_{\rm re}^2 + r^2\ .
\end{flalign}
The observed Compton-scattered flux is then
\begin{flalign}
\label{fluxdustexteqn1}
f_{\e_s^{\rm obs}} & = \frac{\e_s \xi_{\rm dt}L_{\rm disk}\sT\dD^3}{80\pi^3 (2.7\Theta)^2 d_L^2 R_{\rm dt,eff}}
\int^{2\pi}_0 d\phi\ 
\nonumber \\ & \times
\int_{R_{\rm dt,1}}^{R_{\rm dt,2}} \frac{dR_{\rm re}}{x^2} \left(\frac{R}{R_{\rm dt,1}}\right)^{-\zeta}
\nonumber \\ & \times
S_3[2.7\g\Theta(1-\cos\bar{\psi})]N\p_e(\g/\dD)
\end{flalign}
for $\g_{\rm min}<\g$, and
\begin{flalign}
\label{fluxdustexteqn2}
f_{\e_s^{\rm obs}} & = \frac{\e_s \xi_{\rm dt}L_{\rm disk}\sT\dD^3}{80\pi^3 (2.7\Theta)^2 d_L^2 R_{\rm dt,eff}}
\int^{2\pi}_0 d\phi\ 
\nonumber \\ & \times
\int_{R_{\rm dt,1}}^{R_{\rm dt,2}} \frac{dR_{\rm re}}{x^2} \left(\frac{R}{R_{\rm dt,1}}\right)^{-\zeta}
\nonumber \\ & \times
S_3[2.7\g_{\rm min}\Theta(1-\cos\bar{\psi})]
\nonumber \\ & \times
N\p_e(\g_{\rm min}/\dD)
\left(\frac{\g}{\g_{\rm min}}\right)^2\ 
\end{flalign}
for $\g<\g_{\rm min}$.  Modeling by \citet{nenkova08p2} indicated that
$R_{\rm dt,2}/R_{\rm dt,1}=5-10$ and $\zeta=1-2$ with the inner dust
radius $R_{\rm dt,1}$ constrained by Equation (\ref{Rdust}).  A
similar dust torus model to this one was used for Compton
scattering calculations by \citet{sikora13}.

\begin{figure}
\vspace{2.2mm} 
\epsscale{1.0} 
\plotone{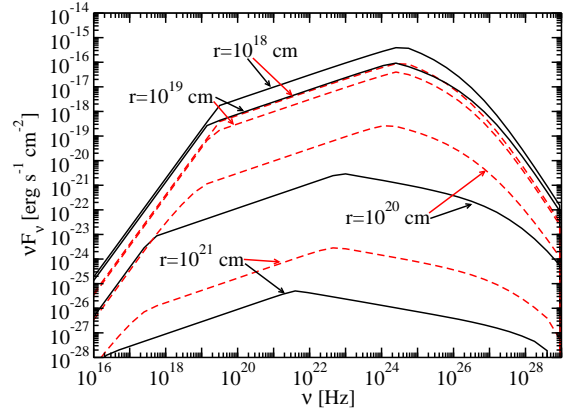}
\caption{The SED of the Compton-scattered dust torus
radiation for a ring geometry (Equation [\ref{fluxdustringeqn1}] and
[\ref{fluxdustringeqn2}]; solid black curves), and for an extended
torus geometry (Equation [\ref{fluxdustexteqn1}] and
[\ref{fluxdustexteqn2}]; dashed red curves) for various distances $r$
from the black hole. }
\label{fluxECdust}
\vspace{2.2mm}
\end{figure}

In Figure \ref{fluxECdust} I plot the Compton-scattered dust radiation
field for the ring and extended dust torus models.  I use the
parameters in Table \ref{compareparamtable}, with the distance of the
emitting region from the black hole ($r$) varying.  At smaller $r$,
the ring model gives a greater Compton-scattered flux.  However, for
$r>R_{\rm dt,1}$ the extended dust torus model gives greater flux,
since in the extended model there are more infrared photons at larger
radii available for scattering.  

\section{Synchrotron Self-Compton}
\label{SSCsection}

The observed $\nu F_\nu$ flux from a spherical plasma blob can
be written in terms of a spherically-symmetric comoving emissivity, 
\begin{flalign}
f_{\e_s^{\rm obs}} = \frac{\dD^4 V\p_b \ep_sj\p(\ep_s)}{4\pi d_L^2}\ .
\end{flalign}

The comoving synchrotron energy density in a spherical blob is
\begin{flalign}
\label{u_sy}
u\p(\ep) = \frac{ 3 d_L^2(1+z)^2f_\e^{sy}}{c^3 t_{\rm v,min}^2 \dD^6 \ep}
\end{flalign}
where 
\begin{flalign}
t_{\rm v,min} = \frac{(1+z)R\p_b}{\dD c}
\end{flalign}
and $R\p_b$ is the comoving blob radius \citep{finke08_SSC}.  The
observed flux from Compton-scattering this radiation field, known as
synchrotron self-Compton (SSC), can be written, using the $\delta$
approximation from Section \ref{isotropic}, as
\begin{flalign}
f_{\e_s^{\rm obs}} & = \frac{3\sT\e_s^{\rm obs}}{32\pi c^2t_{\rm v,min}^2\dD^2}
\int_0^\infty d\ep \frac{f^{sy}_{\e}}{\e^{\prime 4}} 
\nonumber \\ & \times
\frac{N_e\p(\gp)}{\gp}
M_3(4\gp\ep)
\end{flalign}
for $\gp_{\rm min}<\gp$, and
\begin{flalign}
f_{\e_s^{\rm obs}} & = \frac{3\sT\e_s^{\rm obs}}{32\pi c^2t_{\rm v,min}^2\dD^2}
\int_0^\infty d\ep \frac{f^{sy}_{\e}}{\e^{\prime 4}} 
\nonumber \\ & \times
\frac{N_e\p(\gp_{\rm min})}{\gp_{\rm min}}
M_3(2\gp_{\rm min}\ep) \left(\frac{\gp}{\gp_{\rm min}}\right)^{3/2}
\end{flalign}
for $\gp < \gp_{\rm min}$.  The synchrotron flux $f_\e^{sy}$ can be
computed as described by, for example, \citet{finke08_SSC}.

\section{Numerical Results}
\label{numericalsection}

The approximations using the $\delta$-function approximation given in
the previous sections can significantly speed up Compton-scattering
calculations.  It is possible to store the functions $S_0(x)$,
$S_1(x)$, etc. in a data table that is read in at the beginning of the
program, which allows them to be calculated very quickly.  This
essentially eliminates the need to do one of the integrals
numerically.  I have compared calculations using the exact expressions
from, for example, \citet{finke08_SSC} and \citet{dermer09}, with
results using the $\delta$-function approximation, described in
Sections \ref{ECsection} and \ref{SSCsection}.  I present some of
those comparisons here.  As in Sections \ref{disksection},
\ref{BLRsection}, and \ref{dustsection}, I use the baseline model
parameters in Table \ref{compareparamtable}.  In the figures I vary
parameters from this baseline model.

\begin{figure*}
\vspace{10.0mm} 
\epsscale{1.0} 
\plottwo{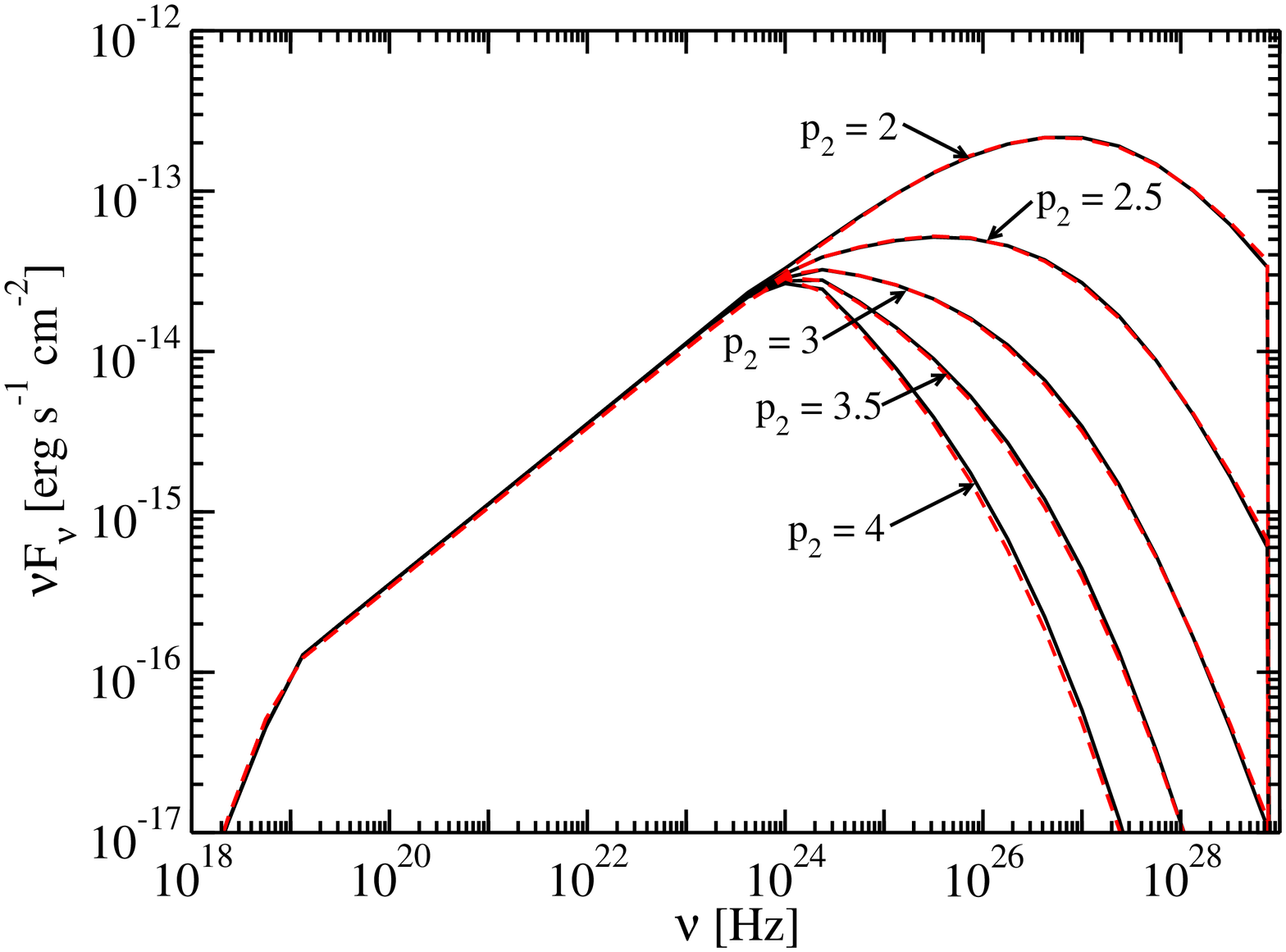}{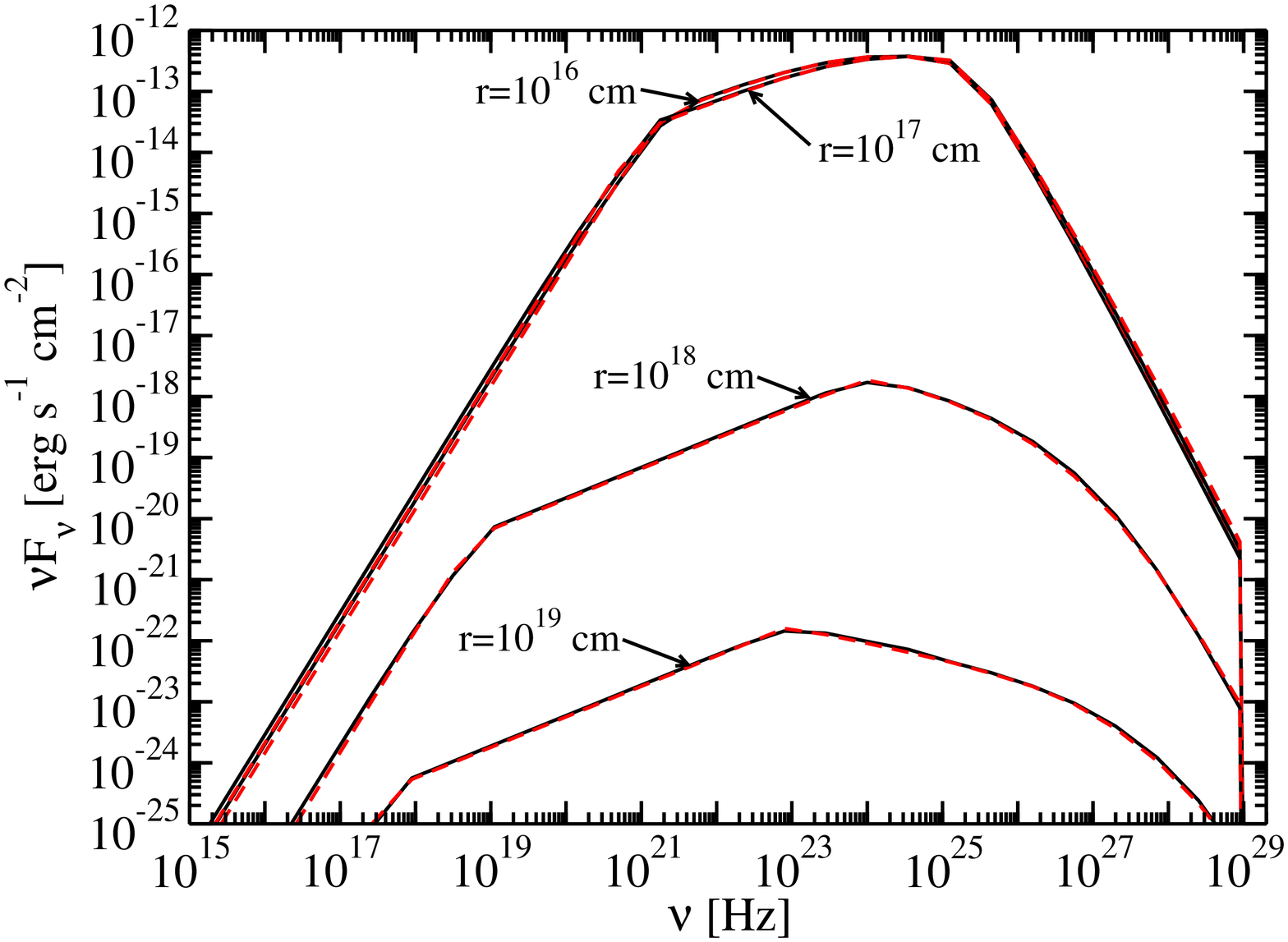}
\caption{Comparison of exact and $\delta$-function Compton-scattering
calculations.  The solid black curve is the exact expression, the
dashed red curve is the $\delta$-function approximation.  {\em Left}:
The scattering of disk photons for different values of $p_2$, as
indicated on the plot.  {\em Right}: The scattering of Ly$\alpha$ BLR
photons assuming a shell geometry for different distances from the
black hole ($r$), as indicated on the plot.  }
\label{fluxcompare}
\vspace{2.2mm}
\end{figure*}

In the left panel of Figure \ref{fluxcompare} I compare Compton
scattering of disk photons, as described in Section \ref{disksection}.
Values of $p_2$ were varied.  The right panel of Figure
\ref{fluxcompare} shows a comparison of scattering of broad line
photons, as described in Section \ref{BLRsection}.  In most cases the
approximation is better than 5\%, although it is higher near the
endpoints or near the frequency associated with $\gp=\gp_b$, where it
can be as high as 12\%.

I also tested the speed of the $\delta$ approximation and compared it
to the full integral calculation.  The exact improvement depends
on many factors, such as the computer used and the numerical
integration technique.  A full exploration of these factors is beyond
the scope of this paper, and not particularly interesting.  For the
calculations I performed, with integrations done by the simple
trapezoid rule, the factor of the speed improvement of the $\delta$
approximation is given in Table \ref{speedup_table}.  The speed-up is
considerable, greater than a factor of 10 in many cases.  Generally
the more integrals a calculation requires, the faster the improvement
in speed.

\begin{deluxetable}{llc}
\tabletypesize{\scriptsize}
\tablecaption{Speed factor improvement of delta approximation.}
\tablewidth{0pt}
\tablehead{
\colhead{Seed Radiation Field} & 
\colhead{Equation} &
\colhead{Speed Improvement} 
}
\startdata
Isotropic Monochromatic & Equation (\ref{u_isomono}) & 3 \\
Point Source Behind Jet & Equation (\ref{u_ptsrc}) & 1.5 \\
Accretion Disk & Equation (\ref{u_disk}) & 26 \\
Broad Line Region (shell) & Equation (\ref{u_BLR}) & 11 \\
Broad Line Region (ring) & Equation (\ref{uBLRring}) & 8 \\
Dust torus (ring) & Equation (\ref{u_dust}) & 8 \\
Dust torus (extended) & Equation (\ref{u_dust_ext}) &  24 \\
\hline
Synchrotron  & Equation (\ref{u_sy}) & 6 
\enddata
\label{speedup_table}
\end{deluxetable}

\section{Beaming Pattern}
\label{beampatternsection}

It is tempting to think that population studies
\citep[e.g.,][]{gardner14} of FSRQs might be able to distinguish
between different geometries for the external scattered photon field.
One can use the results in the previous sections to derive the beaming
pattern of the flux density in the Thomson regime for external Compton
scattering.  This assume a power-law electron distribution,
$N\p_e(\gp) \propto \g^{\prime -p}$.

For an isotropic monochromatic external radiation field, Section
\ref{isomonosection}, the flux density goes as
\begin{flalign}
\label{beam_iso}
F_\nu \propto \nu^{-\alpha}\ \dD^{4+2\alpha}
\end{flalign}
where $\nu=\e_{obs}m_ec^2/h$, $h=6.626\times10^{-27}\ \erg\, \s$ is
Planck's constant, and $\alpha = (p-1)/2$.  Equation (\ref{beam_iso})
was found previously by \citet{georgan01}.

For a point source radially behind the jet, Section \ref{ptsourcesection},  
\begin{flalign}
F_\nu \propto \frac{\nu^{-\alpha}}{r^2}\ \dD^{4+2\alpha}\ (1-\mu_s)^{1+\alpha}\ ,
\end{flalign}
which was found previously by \citet{dermer95}.
For a shell BLR, Section \ref{shellBLRsection}, 
\begin{flalign}
F_\nu & \propto \nu^{-\alpha}\ \dD^{4+2\alpha} \int^{2\pi}_0 d\phi_*\
\int_{-1}^{1} \frac{d\mu_{\rm re}}{x^2}\ 
\nonumber \\ & \times
(1-\cos\bar\psi)^{1+\alpha}\ ,
\end{flalign}
where 
\begin{flalign}
\cos\bar{\psi} = \mu_*\mu_s + \sqrt{1-\mu_*^2}\sqrt{1-\mu_s^2}\cos\phi_*\ ,
\end{flalign}
\begin{flalign}
x^2 = R_{\rm re}^2 + r^2 - 2rR_{\rm re}\mu_{\rm re}\ ,
\end{flalign}
and
\begin{flalign}
\mu_*^2 = 1 - \left(\frac{R_{\rm re}}{x}\right)^2(1-\mu_{\rm re}^2)  \ .  
\end{flalign}
For a ring BLR or ring dust torus, Sections \ref{ringBLRsection} and
\ref{dustsection},
\begin{flalign}
F_\nu \propto \frac{\nu^{-\alpha}}{x^2}\ \dD^{4+2\alpha}\ \int_0^{2\pi} 
d\phi_*\ (1-\cos\bar\psi)^{1+\alpha}\ ,
\end{flalign}
where now
\begin{flalign}
\cos\bar\psi = \mu_s\frac{r}{x} + 
\sqrt{1-\mu_s^2}\sqrt{1-\left(\frac{r}{x}\right)^2}\cos\phi_*\ 
\end{flalign}
and $x^2=R_{\rm re}^2+r^2$.  

\begin{figure*}
\vspace{2.4mm} 
\epsscale{1.0} 
\plottwo{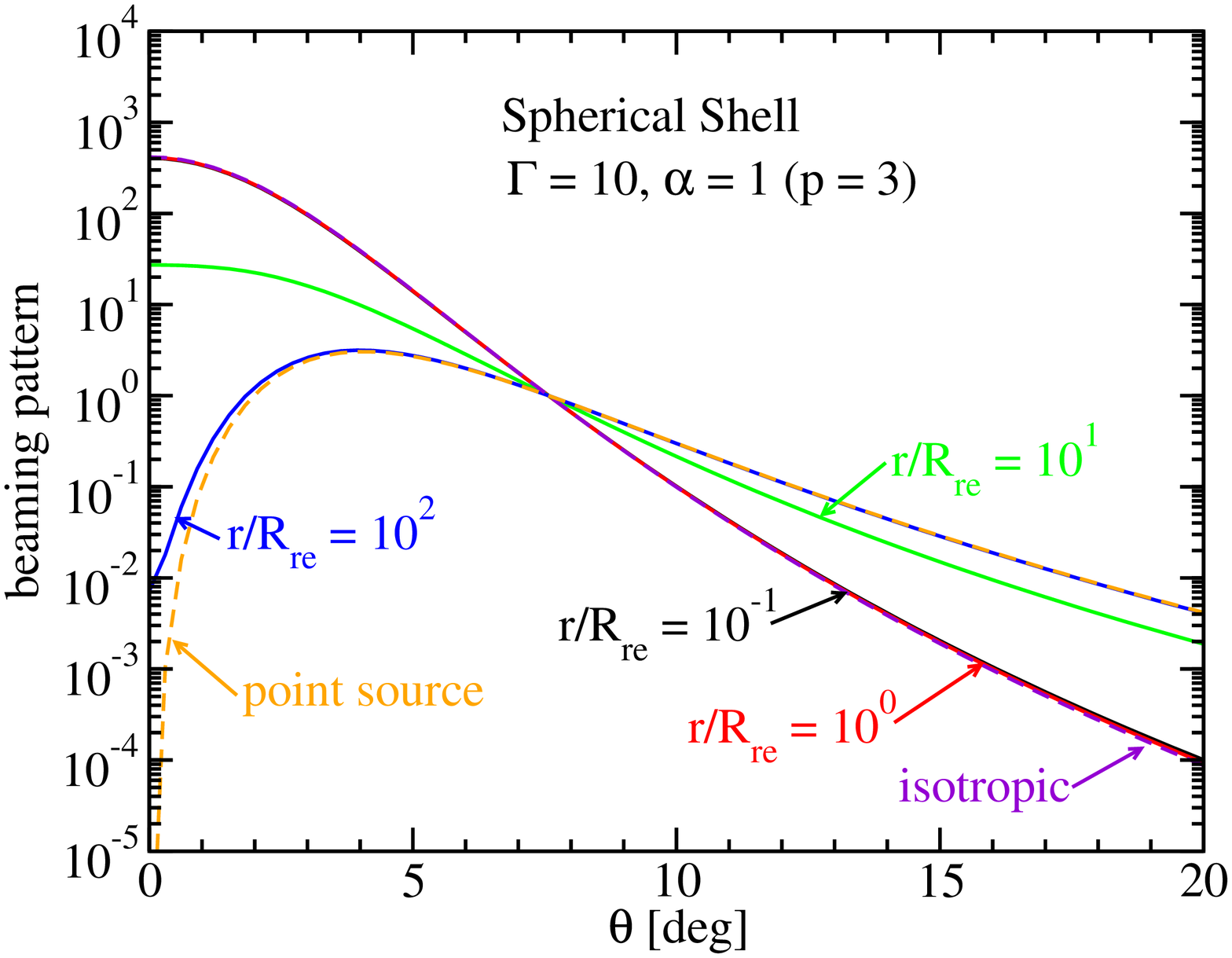}{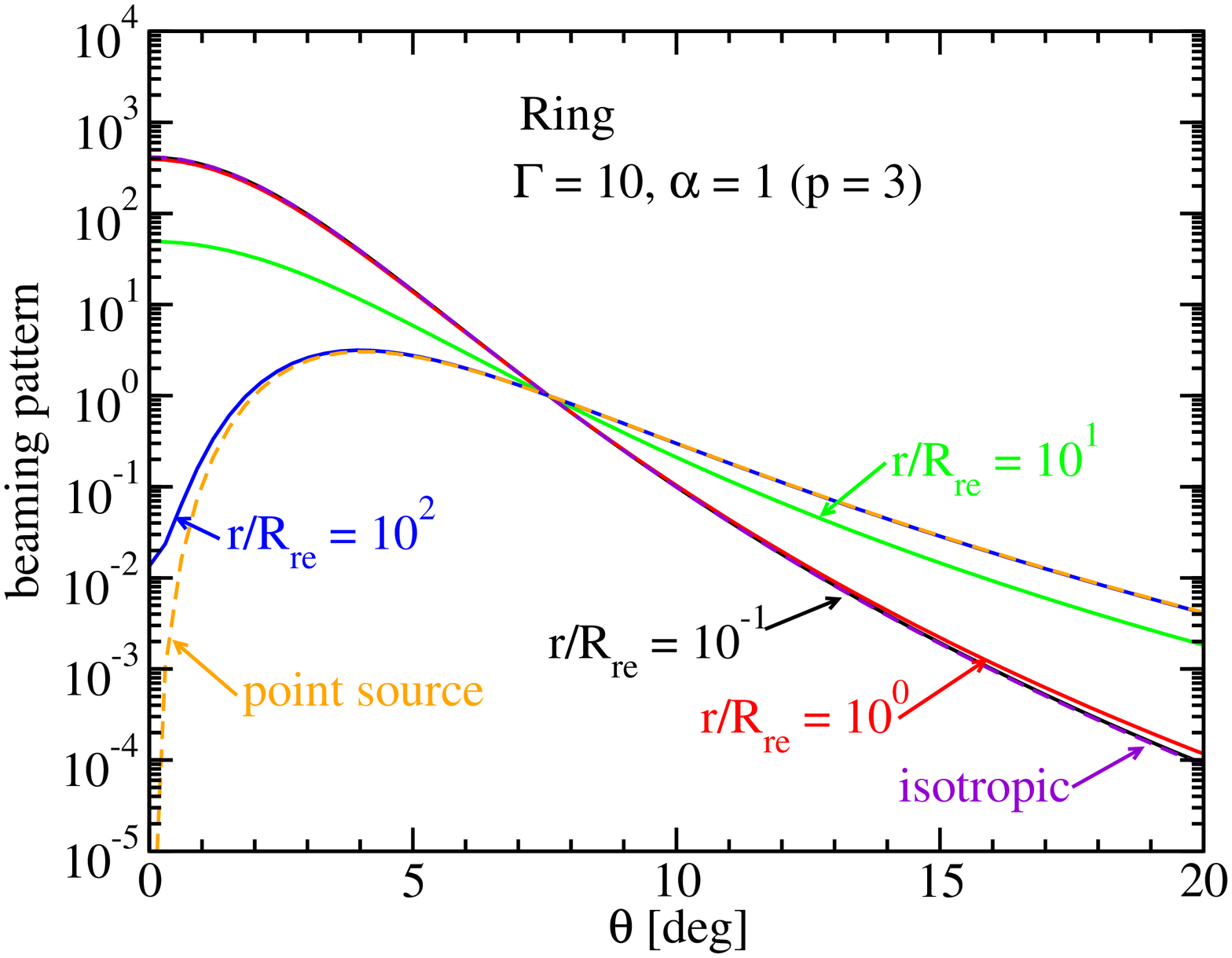}
\caption{The beaming pattern for various external radiation field
geometries, normalized at $\theta=7.5\arcdeg$, for $\G=10$ and $p=3$.
The isotropic and point source cased are shown as the violet and
orange dashed curves, respectively.  The exact cases at various values
of $r/R_{\rm re}$ are shown as the solid curves, as indicated.  {\em
Left}: a spherical shell external radiation field (Section
\ref{shellBLRsection}). {\em Right}: a ring external
radiation field (Section \ref{ringBLRsection}).  }
\label{beam_pattern_fig}
\vspace{2.2mm}
\end{figure*}

For $r \ll R_{\rm re}$ and $r \gg R_{\rm re}$, both the spherical
shell and ring reduce to the isotropic case and point source case,
respectively.  This is demonstrated in Figure \ref{beam_pattern_fig}.
The ring case is actually a small amount brighter at small angles than
the spherical shell for the case where $r = 10 R_{\rm li}$.  When the
emitting region is at this distance, more of the photons from the ring
are at a favorable geometry for efficient Compton scattering compared
to the spherical shell.  However, there is little
difference between the beaming pattern for these different external
radiation field geometries.  It is unlikely that beaming pattern
studies could distinguish between them.

\section{Photoabsorption}
\label{absorbsection}

The same radiation fields that are Compton-scattered to produce $\g$
rays can also absorb the $\g$ rays to produce electron-positron pairs.
The $\g$-ray spectrum is attenuated by a factor
$e^{-\tau_{\g\g}(\e_1)}$ where the absorption optical depth for this
process is
\begin{flalign}
\tau_{\g\g}(\e_1) & = \int_r^\infty d\ell \int_0^{2\pi}d\phi \int_{-1}^{1} d\mu\ 
\nonumber \\ & \times
(1 - \cos\psi) \int_0^\infty d\e\ \frac{u(\e,\Omega; \ell)}{\e m_ec^2} 
\nonumber \\ & \times
\sigma_{\g\g}\left[\frac{\e\e_1(1+z)}{2}(1-\cos\psi)\right]\ ,
\end{flalign}
\citep[e.g.,][]{dermer09}, where $\psi$ is the angle
between the two incoming photons, the $\g\g$ polarization-averaged
absorption cross-section is
\begin{flalign}
\sigma_{\g\g}(s) & = \frac{3}{8}\sT (1-\beta_{\rm cm}^2)
\nonumber \\ & \times
\Biggr[ (3-\beta_{\rm cm}^4)\ln\left(\frac{1+\beta_{\rm cm}}{1-\beta_{\rm cm}}\right) 
\nonumber \\ &
        - 2\beta_{\rm cm}(2-\beta_{\rm cm}^2)\Biggr]\ ,
\end{flalign}
and
\begin{flalign}
\beta_{\rm cm} = \sqrt{1-s^{-1}}
\end{flalign}
\citep[e.g.,][]{gould67,brown73,jauch76,boett14_polar}.  I use these
equations below to find expressions for $\tau_{\g\g}$ by various
external radiation fields.  In so doing I assume the $\g$-ray
photons are traveling directly along the jet axis, i.e., $\mu_s=1$, so
that $\cos\psi \rightarrow \mu$.

\subsection{Absorption by Accretion Disk Photons}
\label{diskabsorb}

For absorption of $\g$ rays by accretion disk photons, using Equation
(\ref{u_disk}),
\begin{flalign}
\tau_{\g\g}(\e_1) & = 10^7\ \frac{\ell_{\rm Edd}^{3/4}M_8^{1/4}}{\eta^{3/4}}
\int_{\tilr}^\infty \frac{d\tilell}{\tilell^2} 
\nonumber \\ & \times
\int_{\tilR_{\rm in}}^{\tilR_{\rm out}}
\frac{d\tilR}{\tilR^{5/4}}\ \frac{\phi(\tilR)}{(1 + \tilR^2/\tilell^2)^{3/2}}
\nonumber \\ & \times
\left[ \frac{\sigma_{\g\g}(\tils)}{\sT}\right] (1- \mu)
\end{flalign}
where $\tilR=R/R_g$, $\tilr = r/R_g$, $\tilell=\ell/R_g$,
\begin{flalign}
\tils = \frac{\e_0(\tilR)\e_1(1+z)(1-\mu)}{2}\ ,
\end{flalign}
and $\mu = (1 + \tilR^2/\tilell^2)^{-1/2}$.  This result was found by
 \citet{dermer09}.  Similar calculations of $\g\g$ absorption by disk
 photons have been done by \citet{becker95} and \citet{donea03}.

\subsection{Absorption by Broad Line Region Photons}
\label{BLRabsorb}

For absorption of $\g$ rays by photons from a broad-line shell,
Equation (\ref{u_BLR}),
\begin{flalign}
\tau_{\g\g}(\e_1) & = 900\ \frac{\xi_{\rm li}\ell_{\rm Edd}}{\e_{\rm li}}
\int_{\tilr}^{\infty} d\tilell
\nonumber \\ & \times
\int_{-1}^{1}\frac{d\mu_{\rm re}}{\tilx^2}
\left[ \frac{\sigma_{\g\g}(\tils)}{\sT}\right](1-\mu_*)
\end{flalign}
where here $\tilr=r/R_g$,  $\tilell=\ell/R_g$,
\begin{flalign}
\mu_*^2 = 1 - \left(\frac{R_{\rm li}}{R_g\tilx}\right)^2(1-\mu_{\rm re}^2)  \ ,
\end{flalign}
\begin{flalign}
\tilx^2 = \frac{ R_{\rm li}^2 + \ell^2 - 2\ell R_{\rm li}\mu_{\rm re}}{R_g^2}\ ,
\end{flalign}
and
\begin{flalign}
\tils = \frac{\e_{\rm li}\e_1(1+z)(1-\mu_*)}{2}\ .
\end{flalign}
Similar calculations of $\g\g$ absorption by BLR photons have been
done by \citet{boett95}, \citet{donea03}, \citet{reimer07}, 
\citet{dermer09}, and \citet{boett16}.

For absorption by photons from a BLR ring, Equation (\ref{uBLRring}), 
\begin{flalign}
\tau_{\g\g}(\e_1) & = 900\ \frac{\xi \ell_{\rm Edd}}{\e_{\rm li}}
\int_{\tilr}^{\infty} \frac{d\tilell }{\tilx^2} 
\nonumber \\ & \times
\left( 1-\frac{\tilell}{\tilx}\right)
\left[ \frac{\sigma_{\g\g}(\tils)}{\sT}\right]
\end{flalign}
where now
\begin{flalign}
\tilx^2 = \left( \frac{R_{\rm li}}{R_g}\right)^2 + \tilell^2\ ,
\end{flalign}
and
\begin{flalign}
\tils = \frac{\e_{\rm li}\e_1(1+z)(1-\tilell/\tilx)}{2}\ .
\end{flalign}

\subsection{Absorption by Dust Torus Photons}
\label{dustabsorb}

For absorption of $\g$ rays by photons from a dust torus, 
using the ring approximation for the dust torus, Equation (\ref{u_dust}), 
\begin{flalign}
\tau_{\g\g}(\e_1) & = 900\ \frac{\xi \ell_{\rm Edd}}{2.7\Theta}
\int_{\tilr}^{\infty} \frac{d\tilell }{\tilx^2} \left( 1-\frac{\tilell}{\tilx}\right)
\nonumber \\ & \times
\left[ \frac{\sigma_{\g\g}(\tils)}{\sT}\right]
\end{flalign}
where here 
\begin{flalign}
\tilx^2 = \left( \frac{R_{\rm dt}}{R_g}\right)^2 + \tilell^2\ ,
\end{flalign}
and
\begin{flalign}
\tils = \frac{2.7\Theta\e_1(1+z)(1-\tilell/\tilx)}{2}\ .
\end{flalign}
Using the extended dust torus model, Equation (\ref{u_dust_ext}), 
\begin{flalign}
\tau_{\g\g}(\e_1) & = 900\ \frac{\xi \ell_{\rm Edd}}{2.7\Theta\tilR_{\rm dt,eff}}
\int_{\tilr}^{\infty} d\tilell
\nonumber \\ & \times
\int_{R_{\rm dt,1}}^{R_{\rm dt,2}}\frac{d\tilde{R_{\rm re}} }{\tilx^2} 
\left( 1-\frac{\tilell}{\tilx}\right)
\left[ \frac{\sigma_{\g\g}(\tils)}{\sT}\right]
\end{flalign}
where now
\begin{flalign}
\tilx^2 = \left( \frac{R_{\rm re}}{R_g}\right)^2 + \tilell^2\ 
\end{flalign}
and $\tilR_{\rm dt,eff}=R_{\rm dt,eff}/R_g$.

Similar calculations of $\g\g$ absorption by interactions with dust
torus photons have been done by \citet{protheroe97} and
\citet{donea03}.

\subsection{The Escape of Gamma Rays}

\begin{figure*}
\vspace{10.0mm} 
\epsscale{1.0} 
\plottwo{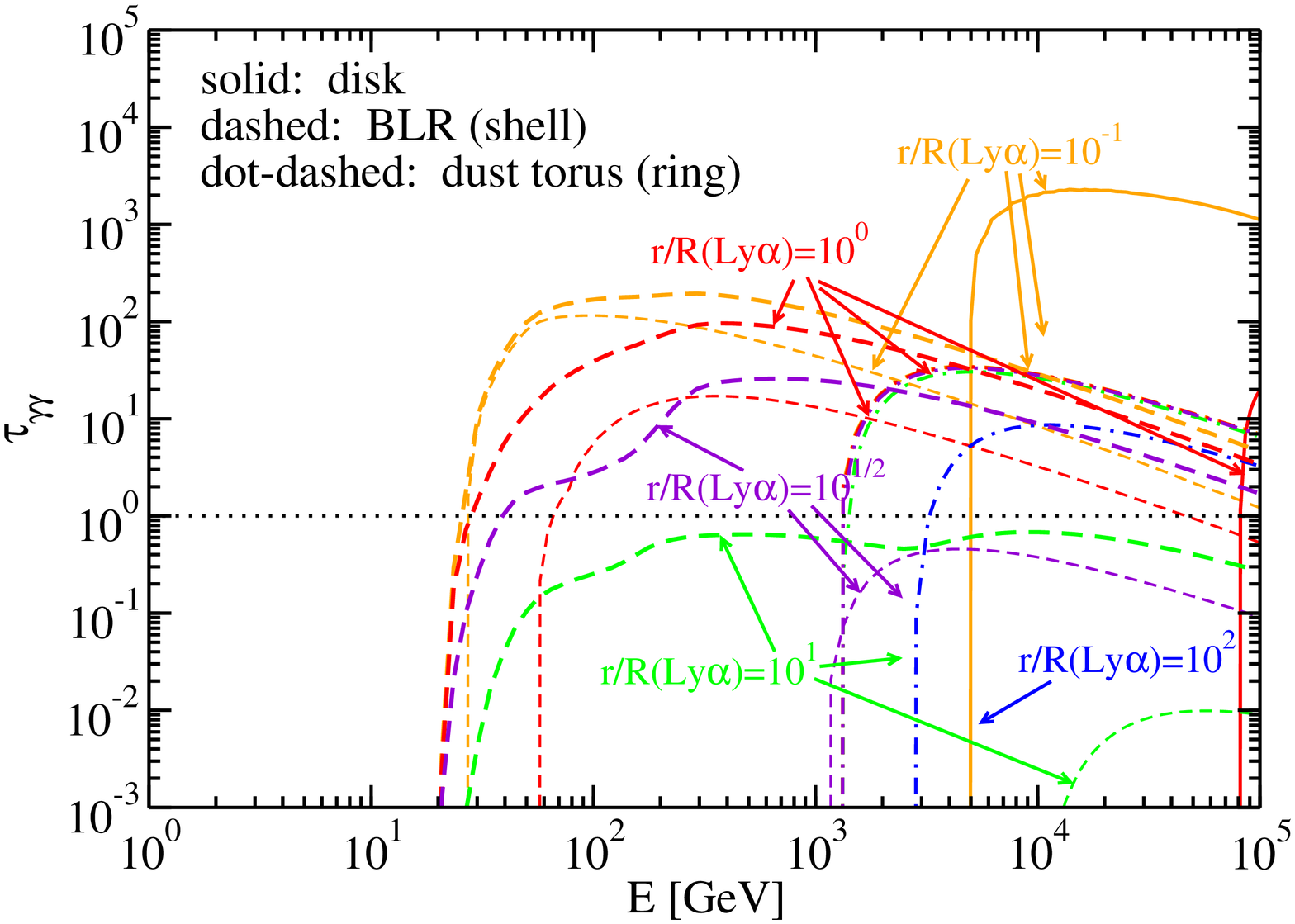}{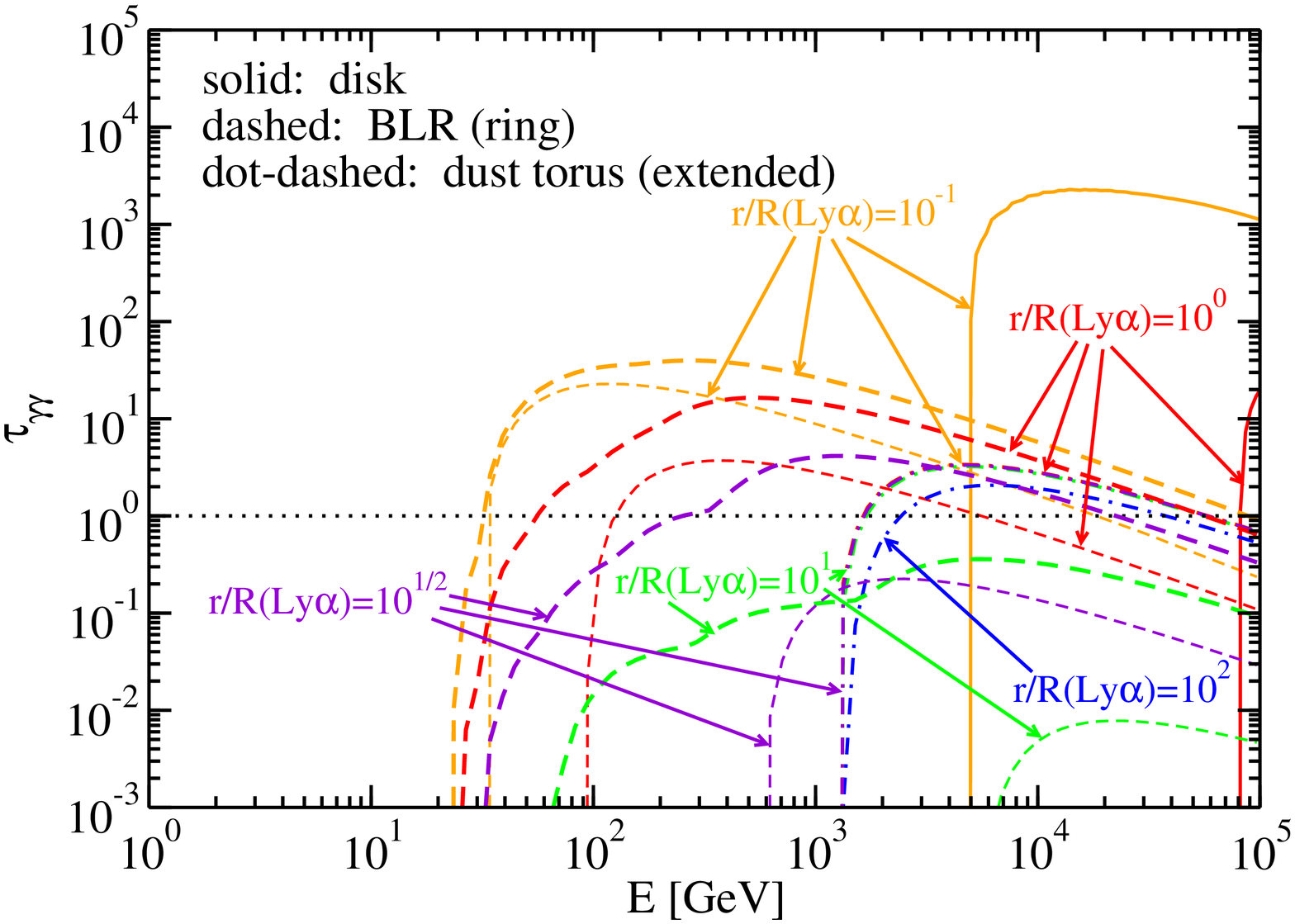}
\caption{The $\g\g$ absorption optical depth from accretion disk
photons (solid), BLR photons (dashed), and dust torus photons
(dot-dashed).  Absorption from all of the BLR line emission is given
by the thick dashed curves, while absorption by only Ly$\alpha$ is
given by the thin dashed curves.  The dotted lines indicate
$\tau_{\g\g}=1$.  {\em Left}: spherical shell BLR geometry and ring
dust torus geometry.  {\em Right}: ring BLR geometry and extended dust
torus geometry. }

\label{tau_ext_fig}
\vspace{2.2mm}
\end{figure*}

In Figure \ref{tau_ext_fig} is plotted $\tau_{\g\g}$ versus photon
energy $E$ for photons from the accretion disk (Section
\ref{diskabsorb}), BLR (Section \ref{BLRabsorb}), and dust torus
(Section \ref{dustabsorb}), for $\g$ rays originating at various
distances $r$ from the black hole.  The BLR parameters are chosen to
be consistent with 3C 454.3, as described in Appendix \ref{BLRmodel}.
The accretion disk and dust torus parameters are given in Table
\ref{absorbparamtable}.  The black hole mass is consistent with that
found for 3C 454.3 by \citet[][see the discussion therein]{bonnoli11},
and the disk luminosity was chosen so that $L_{\rm disk}=10L(5100\ {\rm
\AA})$.  Dust torus parameters are consistent with Equation
(\ref{Rdust}).  Note that energy ($E$) is in the frame of the galaxy,
so that the observer sees absorption at energies a factor $(1+z)$
lower and that distances are given in terms of $R({\rm
Ly}\alpha)=1.1\times10^{17}\ \cm$.

\begin{deluxetable}{lcc}
\tabletypesize{\scriptsize}
\tablecaption{Accretion Disk and Dust Torus Parameters}
\tablewidth{0pt}
\tablehead{
\colhead{Parameter} &
\colhead{Symbol} &
\colhead{Value}
}

\startdata
Black hole mass [$M_{\odot}$]      &    $M_{\rm BH}$    & $1.2\times10^9$ \\
Gravitational radius [cm]          &    R$_g$           & $1.8\times10^{14}$ \\
Disk luminosity [$\erg\ \s^{-1}$]  &    $L_{\rm disk}$  & $2\times10^{46}$  \\
Accretion efficiency               &    $\eta$          & $1/12$ \\
Inner disk radius [$R_g$]          &    $R_{\rm in}$    & $6$        \\
Outer disk radius [$R_g$]          &    $R_{\rm out}$   & $200$  \\
Dust torus temperature [K]         &    $T_{\rm dt}$    & $10^3$ \\
Dust torus scattering fraction     &    $\xi_{\rm dt}$  & $0.1$ \\
Dust torus (inner) radius [cm]     &    $R_{\rm dt,1}$  & $3.5\times10^{18}$ \\
Dust torus (outer) radius [cm]     &    $R_{\rm dt,2}$  & $3.5\times10^{19}$ \\
Dust torus cloud density gradient  &    $\zeta$         & 1.0 
\enddata
\label{absorbparamtable}
\end{deluxetable}

Absorption by disk photons does not play a large role for the
parameters explored here; they are important for $>5$\ TeV for the
lowest value of $r$ shown, where photons have not escaped anyway due
to absorption by BLR and dust torus photons.

For $r<R({\rm Ly}\alpha)$, BLR absorption from Ly$\alpha$ dominates,
although absorption by other lines is not negligible.  At $r>R({\rm
Ly}\alpha)$, other lines with larger radii dominate the BLR
absorption.  The flattened, ring geometry for the BLR generally has
lower $\tau_{\g\g}$, and requires a higher energy before pair
production is important, as pointed out previously by
\citet{stern14}.  For distances $r>10R({\rm Ly}\alpha)$,
$\tau_{\g\g}<1$ from BLR photons, regardless of the geometry.  Based
on Figure \ref{tau_ext_fig}, the detection of $\approx 60$\ GeV
photons from 3C 454.3 \citep{pacciani14} produced at $z=0.859$
restricts their production to $r\ga 10^{1} R(\rm{Ly}\alpha) \approx
1.1\times10^{18}\ \cm$ for a shell BLR geometry and $r\ga 10^{1/2}
R(\rm{Ly}\alpha) \approx 3.4\times10^{17}\ \cm$ for a ring BLR
geometry.  Similar constraints can be made for other sources.  Cutoffs
in $\g$-ray spectra of FSRQs by BLR photons may have already been
detected by the {\em Fermi}-LAT \citep{stern14}.  Those authors assume
the cutoffs are due to absorption by Lyman continuum photons from the
BLR (13.6 eV photons), which I do not consider since this feature is
not included in the template by \citet{vandenberk01}; see Appendix
\ref{BLRmodel}.  The absorption of $\ga 10\ \GeV$\ $\g$ rays by BLR
photons can have implications for using FSRQs to constrain the
extragalactic background light \citep{reimer07}.

Figure \ref{tau_ext_fig} shows that absorption by dust torus photons
is mostly constant with $r$ for $r \ll R_{\rm dt}\sim 10^2 R({\rm
Ly}\alpha)$.  Above this value of $r$, the absorption by dust torus
photons decreases for both the ring and extended geometries.  The
opacity is lower for the ring geometry, and pair production does not
occur until slightly higher energies.  If $\g$ rays are produced at $r
\ll R_{\rm dt}$, one does not expect to detect $\g$ rays at $\ga 1.3$\
TeV for either the ring geometry or extended dust geometry, again, in
the source frame.  Indeed no $\g$ rays have been detected at these
energies from an FSRQ, despite several FSRQs having been detected by
imaging atmospheric Cherenkov telescopes.  This includes 3C 279
\citep{albert08_3c279}, 4C 21.35 \citep{aleksic11}, PKS 1510$-$089
\citep{abramowski13_pks1510}, and PKS 1441+25
\citep{abeysekara15_pks1441, ahnen15_pks1441}.  It has been announced
that the FSRQ S4 0954+65 has been detected by MAGIC
\citep{mirzoyan15}, although the maximum energy at which it was
detected is not yet public knowledge.  These objects may have minor
differences in parameters compared to those used to generate Figure
\ref{tau_ext_fig}, but differences are probably not too large.

\section{Location of the Gamma-Ray Emitting Region}
\label{signaturesection}

\begin{figure}
\vspace{10.0mm} 
\epsscale{1.0} 
\plotone{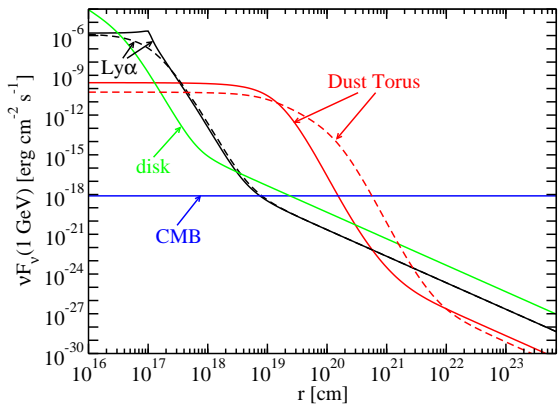}
\caption{Plot of the $\nu F_\nu$ flux at 1 GeV (in the observer's
frame) from Compton-scattering of various external radiation fields as
a function of distance of the emitting region from the black hole.
The seed photon radiation field is indicated on the plot.  The solid
black curve indicates a shell BLR geometry, while the dashed lack
curve indicates a ring BLR geometry. The solid red curve indicates the
ring dust torus geometry, while the dashed red curve indicates the
extended dust torus geometry.}
\label{fluxdist}
\vspace{2.2mm}
\end{figure}

The dominant seed photon source for Compton scattering depends on
the distance of the emitting region from the black hole.  The EC flux
at 1 GeV (in the observer's frame) as a function of the emitting
region's distance from the black hole is plotted in Figure
\ref{fluxdist}.  For this plot, the accretion disk and dust torus
parameters are given in Table \ref{absorbparamtable}.  I plot only the
Ly$\alpha$ line with parameters as described in Appendix
\ref{BLRmodel}, although other lines contribute.  Other
parameters are taken from the modeling of 3C 454.3\footnote{This
source has a redshift $z=0.859$ giving it a luminosity distance
$d_L=5.5$\ Gpc assuming a cosmology with $(h,\Omega_m, \Omega_\Lambda)
= (0.7, 0.3, 0.7)$.} by \citet{cerruti13_3c454} in 2010 November,
which they referred to as ``epoch B''.  These parameters can be seen
in Table \ref{jetparamtable}.  These authors use a (co-moving frame)
log-parabola electron distribution described by
\begin{flalign}
N\p_e(\gp) = N\p_e(\gp_{pk})
\left(\frac{\gp}{\gp_{pk}}\right)^{-[2+b\log_{10}(\gp/\gp_{pk})]}\ ,
\end{flalign}
and consequently, so do I in this section.  Note that in Table
\ref{jetparamtable}, unlike \citet{cerruti13_3c454}, $N\p_e(\gp_{pk})$
is in terms of absolute number of electrons rather than electron
density.  The only parameter that changes with time is $r$.

In Figure \ref{fluxdist}, one can see which radiation fields 
dominate the scattering process at different radii.  At the lowest
radii, scattering of disk radiation dominate.  At $r\ga
3\times10^{16}\ \cm$, scattering of the Ly$\alpha$ broad line 
begins to dominate.  At $r\la R({\rm Ly}\alpha)$ the emission from
scattering Ly$\alpha$ line radiation is approximately constant,
although it decreases somewhat for the ring geometry.  At $r\approx
R({\rm Ly}\alpha)$ there is an enhancement in scattering in the
shell geometry, due to the proximity of the seed photons to the
emitting region in this geometry, that is not present in the ring
geometry.  This enhancement was noted previously by several authors,
\citep[e.g.,][]{ghisellini96,donea03}.  For $r\ga R({\rm Ly}\alpha)$,
the EC-Ly$\alpha$ flux decreases extremely rapidly, as $f_\e \propto
r^{-7.7}$ until $r\approx 10^2R({\rm Ly}\alpha)$ for both the shell
and ring geometries.  Above this radius, the BLR appears as a
point source behind the jet, so that the flux decreases as $f_\e
\propto r^{-2}$.  The EC-dust torus flux follows the same pattern for
the ring geometry, as expected, except at larger radii.  That is, it
is $f_\e \propto r^0$ for $r\la R_{\rm dt}$; $f_\e \propto
r^{-7.7}$ for $R_{\rm dt}\la r\la 10^2 R_{\rm dt}$; and $f_\e \propto
r^{-2}$ for $10^2 R_{\rm dt}\la r$.  For the extended dust torus
geometry, however, $f_\e\propto \e^{-2.5}$, between $R_{\rm dt,1}$ and
$R_{\rm dt,2}$, decreasing more slowly than the ring geometry at these
radii.  With this dust torus model, Compton-scattering of dust torus
radiation to produce $\g$ rays is more viable at higher radii than the
ring model.  The scattering of the disk radiation field as a function
of radius has been described previously by \citet{dermer02}.  One
consequence of this is that, at large radii ($r\ga 3\times10^{20}\ \cm
\approx 100\ \pc$) the scattering of disk radiation actually
dominates over scattering of Ly$\alpha$ line and dust torus radiation.
However, this has little practical effect, as at these radii the
scattering of diffuse CMB photons dominates over all other components
shown on this plot.

\begin{deluxetable}{lcc}
\tabletypesize{\scriptsize}
\tablecaption{Jet Parameters}
\tablewidth{0pt}
\tablehead{
\colhead{Parameter} &
\colhead{Symbol} &
\colhead{Value}
}
\startdata
Lorentz factor               &    $\G$                & $40$ \\
Doppler factor               &    $\dD$               & $40$ \\
Magnetic Field [G]           &    $B$                 & $0.56$  \\
ED\tablenotemark{a}\ peak Lorentz factor    &    $\gp_{\rm pk}$      & $180$ \\
ED\tablenotemark{a}\ normalization          &    $N\p_e(\gp_{pk})$    & $2.4\times10^{50}$        \\
ED\tablenotemark{a}\ width parameter        &    $b$                 & $1$  \\
\enddata
\tablenotetext{a}{Electron Distribution}
\label{jetparamtable}

\end{deluxetable}

For FSRQs, one expects that the nonthermal optical synchrotron and EC
$\g$-ray emission is produced by approximately the same electrons.
There is ample evidence of this in the correlation between optical and
$\g$-ray variability in these objects
\citep[e.g.][]{chatterjee12,cohen14,ramak16}.  An emitting region
producing optical and $\g$ rays is expected to be moving at high
relativistic speeds ($\G \gg 1$).  If the emission originates from a
region deep within the BLR, it quickly travels out of it, with
external radiation fields becoming less geometrically favored for
Compton scattering.  Since the optical and $\g$-ray photons are
produced by the same electrons, the observed $\g$ rays should decrease
relative to the optical.  I compute the ratio of integrated $\g$-ray
flux from EC processes to the optical ($R$ band) flux density from
synchrotron emission.  For a blob traveling at constant velocity along
the jet, its distance from the base of the jet at time to the observer
$t_{\rm obs}$ is given by
\begin{flalign}
\label{robs}
r = r_0 + \frac{\beta c \dD \G\ (t_{\rm obs}-t_{\rm obs,0})}{1+z}
\end{flalign}
assuming it is at a distance $r_0$ at $t_{\rm obs,0}$.  Included are
Compton scattering of accretion disk, BLR, and dust torus radiation.
I include scattering of all of the lines described in Appendix
\ref{BLRmodel}.  Since the calculations are restricted to $< 10$\ GeV,
$\g\g$ absorption does not play a role, since there does not seem to
be any absorption at $<10$\ GeV in FSRQs (Section
\ref{absorbsection}).  Synchrotron emission is calculated using
formulae found in \citet{crusius86} and \citet{finke08_SSC}.

\begin{figure}
\vspace{2.2mm} 
\epsscale{1.0} 
\plotone{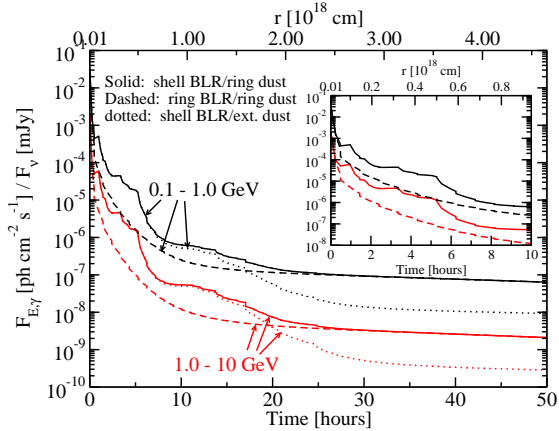}
\caption{Predicted ratio of integrated $\g$-ray flux to $R$ band flux
density for a moving blob as a function of observed time.  The energy
range for the integrated $\g$-ray flux is indicated on the plot.
Solid curves show results for a spherical shell BLR and ring dust
torus; dashed curves show results for a ring BLR and ring dust torus;
the dotted curves show results for a shell BLR and extended dust
torus.  The labels at the top display the distance of the blob from
the base of the jet, $r$, at a given observer time.  The inset shows
the first 10 hours in more detail.}
\label{gammaopt_ratio}
\vspace{2.5mm}
\end{figure}

Results of this calculation can be seen in Figure
\ref{gammaopt_ratio}.  At first, the scattering of the accretion disk
radiation dominates the $\g$-ray emission, but it quickly becomes very
faint, within the first $\approx 0.25$ hour.  In the spherical shell
geometry, as the blob approaches the radius of a particular line ($r
\approx R_{\rm li}$) there is a peak in the flux, since there is
essentially 0 distance between the blob and the photons source.  This
can be seen at $r\approx R({\rm Ly}\alpha)\approx 10^{17}\ \cm$.  It
can also be seen at later times where it approaches other shells,
e.g., $r\approx 3\times10^{17}\ \cm$ where it approaches the shell for
the lines from \ion{C}{3}, \ion{N}{3}, \ion{Si}{4}, \ion{O}{4}], and
especially \ion{C}{4}.  There is another peak at $r\approx
5\times10^{17}\ \cm$ where it approaches the shell for the ions for
Ly$\beta$ and \ion{O}{6}.  These peaks are not seen for the ring
geometry, since with this geometry the blob does not approach the
physical location where the lines are emitted.  In general, the ring
geometry results in less emission than the shell geometry.  Comparing
the decay for the ring dust torus model and extended dust torus model,
clearly the extended dust torus model has lower flux, although it
decays less rapidly, consistent with the results shown in Figure
\ref{fluxdist}.

\begin{figure}
\vspace{2.2mm} 
\epsscale{1.0} 
\plotone{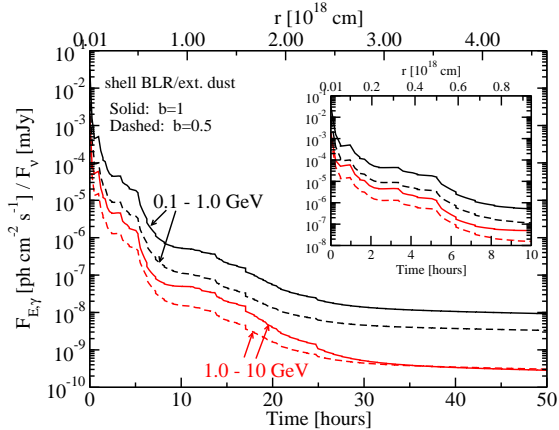}
\caption{Same as Figure \ref{gammaopt_ratio}, for a
shell BLR and extended dust torus model, but now plotting for
different values of the log-parabola electron distribution parameter
$b$.}
\label{gammaopt_ratio_spectrum}
\vspace{2.5mm}
\end{figure}

As the blob travels along the jet, the electron distribution will
likely change.  If the normalization but not the shape of the electron
distribution changes (and the thermal accretion disk is not an
important emission component), the ratio of $\g$-ray to optical flux
as a function of time should follow the basic shape shown Figure
\ref{gammaopt_ratio}.  However, if the shape of the electron
distribution changes, there could be additional variation in this
ratio.  In Figure \ref{gammaopt_ratio_spectrum} I plot the $\g$-ray
flux to optical flux density ratio for the shell BLR and extended dust
torus model for the same set of parameters (Table
\ref{jetparamtable}), but for both $b=1.0$ and $b=0.5$.  The ratio
clearly changes in response to changes in the electron distribution.
However, changes due to this are small compared to the overall changes
due to the falling external radiation field, where the ratio falls by
many orders of magnitude within one day.

\begin{figure}
\vspace{2.5mm} 
\epsscale{1.0} 
\plotone{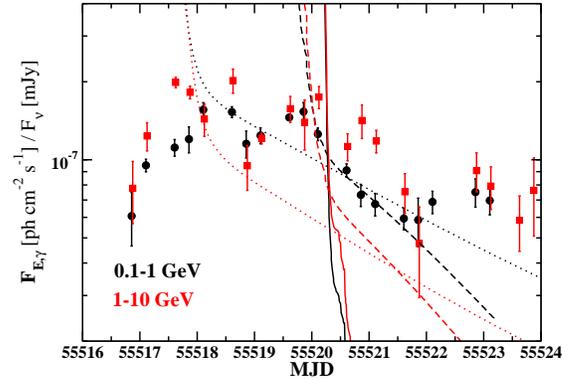}
\caption{Ratio of integrated $\g$-ray flux to $R$ band flux density
observed during a flare from 3C 454.3 in 2010 November.  The curves
show the predictions from Figure \ref{gammaopt_ratio}.  The energy
range for the integrated $\g$-ray flux is indicated on the plot.
Solid curves show results for a spherical shell BLR and ring dust
torus.  The dashed curves show this result adjusted to match the
0.1-1.0 GeV observations between MJD 55520 and MJD 55522.  The dotted
curves show the extended dust torus curve from Figure
\ref{gammaopt_ratio} adjusted to match the 0.1-1.0 GeV observations in
the same time period.  The 0.1-1 GeV data and curves have been
multiplied by a factor 0.05.  }
\label{gammaopt_ratio_data}
\vspace{2.2mm}
\end{figure}

In Figure \ref{gammaopt_ratio_data} these theoretical curves are
plotted with the $\g$-ray to optical flux ratio as a function of time
for the 2010 November flare in 3C 454.3.  The LAT $\g$-ray data were
previously presented by \citet{abdo11_3c454.3} and the Whole Earth
Blazar Telescope (WEBT) optical data by \citet{vercellone11}.  These
$R$ band optical data were corrected for Galactic extinction using the
model of \citet{schlafly11}, as found on the NASA Extragalactic
Database\footnote{\url{https://ned.ipac.caltech.edu}}.  The $\g$-ray
time bins were 6 hours, and for most of these bins there were multiple
WEBT data points, so in each 6-hour LAT time bin the WEBT data were
averaged.  Several of the LAT 6-hour bins did not have any optical
data points in them, in which case the bin is not plotted.  The flares
were quite bright in both $\g$ rays and optical; the latter indicates
that contamination to the optical during the flare is likely to be
minimal. The solid curves show the curves from Figure
\ref{gammaopt_ratio_data} with a shell BLR and ring dust torus
adjusted so that $t=0$ corresponds to MJD 55520.0.  The blob rapidly
moves out of the BLR so that these curves decay too quickly to explain
the flare.  If the flare occurs inside the BLR, it must be the result
of a superposition of many different flaring regions, rather than a
single region.  A $\g$-ray sub-flare began on $\approx$\ MJD 55519.0
that lasted until $\approx$\ MJD 55522 \citep{abdo11_3c454.3}.
Although this flare is too long to be explained by scattering of BLR
photons in a single relativistically moving component, it may be
explained by scattering of dust torus photons.  Hence I also show the
model adjusted to reproduce the ratio of the 0.1-1.0 GeV flux to the
$R$ band flux for the time period where the ratio is decaying, i.e.,
MJD 55520-MJD 55522 (the dashed black curve).  In this case, the decay
phase is described by the scattering of dust torus radiation.  Once
can see that the curve can indeed reproduce the decaying portion of
this flare in the 0.1-1.0 GeV range.  However, when the ratio of the
1.0-10.0 GeV flux to $R$ band flux is adjusted by the same amount (red
dashed curve) the curve clearly is not able to reproduce the
observations.  The $\g$-ray spectrum hardens during this decay phase,
as noted by \citet{abdo11_3c454.3}.  The models with the extended dust
torus are shown as the dotted curves, again adjusted to reproduce the
ratio of 0.1-1.0 GeV flux to $R$ band flux density observations in the
entire MJD 55519-MJD 55522 range.  The scattering of dust torus
photons can reproduce the ratio of the 0.1-1.0 GeV flux, except for
during MJD 55519.5-MJD 55520.0.  Again, it does not also reproduce the
1.0-10.0 GeV curve.

My purpose here is not to reproduce the observed curves exactly, but
to demonstrate that Compton scattering of dust torus photons by a
single emitting jet component can in principle reproduce the entire
sub-flare, from MJD 55519 to 55520, while scattering of disk or BLR
photons by a single jet component cannot.  Discrepancies between the
dotted curves and the data could be explained by changes in other
physical parameters in the emitting region, such as the magnetic field
or electron distribution parameters.  Simple ``one-zone'' models
where $\g$ rays are created by Compton scattering of disk or BLR
photons seem to be ruled out.  More generally, EC models scattering
dust torus photons with the extended dust torus model with $R_{\rm
dt,2}=10R_{\rm dt,1}$, as suggested by
\citet{nenkova08p1,nenkova08p2}, should be viable for flares occurring
at distances
\begin{flalign}
r \la R_{\rm dt,2} \approx 11\ L_{45}^{1/2}\ T_{\rm dt,3}^{-2.6}\ \pc
\end{flalign}
or lasting 
\begin{flalign}
\Delta t \la 14\ \delta_{1.5}^{-2} \left[\frac{\delta_D}{\Gamma}\right]\ 
L_{45}^{1/2}\ T_{\rm dt,3}^{-2.6}\  (1+z)\ \dday\ 
\end{flalign}
where $\delta_{1.5}=\dD/30$ and $L_{45}=L_{\rm disk}/(10^{45}\ \erg\
\s^{-1})$.  This is compatible with evidence that flares occur at
these distances from the black hole in many FSRQs and BL Lacs.  This
includes $\g$-ray flares in 3C 279 at $r\approx 5$\ pc
\citep{abdo10_3c279}, $r\approx17$\ pc PKS 1510$-$089
\citep{marscher10}, $r\approx14$\ pc in OJ 287 \citep{agudo11_oj287},
$r\approx 12$\ pc in AO 0235+164 \citep{agudo11_ao0235}, and
$r\approx10$\ pc from PKS 1510$-$089 \citep{orienti13}.  These flares
could occur at the location of the 7 mm (43 GHz) core and still have
$\g$-ray emission from Compton-scattering of dust photons, although if
they occurred at much larger $r$, this scenario would not be viable.  {An 

However, at these larger distances, the cooling timescale from Compton
scattering of dust torus emission might be too long to explain the
observed rapid decay in $\g$-ray flares.  The energy density of photons
from the dust torus can be estimated as
\begin{flalign}
\up_{dt} & = \chi\ \frac{\G^2 \xi_{\rm dt}L_{\rm disk}}{4\pi R_{\rm dt,1}^2c} 
\nonumber \\ &
= 2.3\times10^{-5}\ \chi\ \xi_{\rm dt,-1}\ T_{\rm dt,3}\
\erg\ \cm^{-3}
\end{flalign}
where $\chi$ is a geometric factor taking into account the dust torus
geometry and $\xi_{\rm dt,-1}=\xi_{\rm dt}/0.1$.  The parameter
$\chi=1$ for the ring geometry with $r\ll R_{\rm dt}$.  One can use
Figure \ref{fluxdist} to estimate $\chi=0.01$ for the extended dust
torus at $r=10$\ pc.  Assuming $T_{\rm dt}\le 2000$\ K, the observer
frame cooling timescale in the Thomson regime can be estimated at
$r=10$\ pc as
\begin{flalign}
t_{\rm cool} & = \frac{1+z}{\dD}\frac{3m_ec^2}{4c\sT \up_{\rm dt}\gp}
\nonumber \\ &
\ge 36\ (1+z)^{1/2}\ \chi_{-2}\ 
\delta_{1.5}^{-2}\ 
\left(\frac{\dD}{\G}\right)^2 
\nonumber \\ & \times
\left( \frac{m_ec^2\e}{\GeV}\right)^{-1/2}\
\hour\ 
\end{flalign}
where $\chi_{-2}=\chi/0.01$.  \citet{abdo11_3c454.3} find a falling
timescale of $15\pm2$\ hour for the $\g$-ray flare centered on MJD
55520.  This would be marginally consistent with this cooling
timescale for particularly large Doppler factors.  On the other hand,
the $\g$-ray flare from PKS 1510$-$089 around MJD 55850 has a falling
timescale of $1.2\pm0.15$\ hour \citep{brown13}, which would be a
problem if it was produced at 10 pc from the black hole, as suggested
by \citet{orienti13}.  It might be possible to explain this rapid
cooling by adiabatic losses.


\section{Summary}
\label{discussion}

The location of the $\g$-ray emitting region is blazar jets is
poorly known; modeling EC jet emission is necessary to determine its
origin, but doing this correctly requires the full angle-dependent
Compton cross-section.  Consequently, I have created a novel
$\delta$-function approximation for Compton scattering that is fully
angle dependent, valid in both the Thomson and extreme Klein-Nishina
regimes.  It assumes that all of the scattered photons have the same
energy of the mean scattered photon.  It allows one to eliminate an
integral in Compton scattering calculations, so that in applications
for scattering in blazar jets, it is numerically faster by up to
approximately a factor of 10.  It is likely to have applications to
other high-energy astrophysical situations where accurate Compton
scattering calculations are important.

I have used this approximation to derive formulae for Compton
scattering of external radiation fields of interest to blazars;
namely, external radiation originating from the accretion disk, BLR,
and dust torus.  The Compton-scattered accretion disk formula is
essentially the same as the one from \citet{dermer09}, only with the
new approximation.  Formulae for Compton-scattering of BLR photons are
provided for two geometries, considering the line emission as
originating from spherical shell and from a flat annulus.  I have also
provided a method for estimating the relative radii and luminosities
of the various broad lines that make up the BLR (Appendix
\ref{BLRmodel}).  I also provide formulae for computing the scattering
of dust torus photons using two dust torus models, one as a thin
annulus, and one as an extended torus made up of many dust clouds.
The latter is similar to the one described by \citet{sikora13},
although I take into account the full angle dependence in the
Compton-scattering calculation.  I derive beaming patterns for
scattering of spherical shell external radiation fields or annulus
external radiation fields in the Thomson regime, the latter of which
could represent either a broad emission line or a dust torus.  I show
that if the shell or annulus has a radius $R_{\rm re}$, then for
distances $r\ll R_{\rm re}$, the beaming pattern resembles that of an
isotropic radiation field; and for distances $r\gg R_{\rm re}$, it
resembles the beaming pattern of a point source behind the jet.  
The beaming pattern for the spherical shell and annulus seed radiation
field geometries are similar enough to each other that beaming pattern
studies are unlikely to distinguish between them.

Using my models for radiation from the BLR and dust torus, I have
computed the $\g\g$ absorption optical depth as a function of $\g$-ray
energy and distance from the black hole.  I believe these are the most
detailed of this sort of calculation that has yet been performed.
This calculation shows that photons below $\approx20\ \GeV$ in the
source frame are always able to escape unabsorbed.  For $\g$ rays
emitted $r \ga 10R({\rm Ly}\alpha)$, $\tau_{\g\g}<1$ everywhere, at
least for the fiducial parameters I used.  For emission of $\g$ rays
in between these radii, the escape of $\g$ rays is dependent on the
details of the BLR structure, which I have realistically computed.
The flattened, ring geometry lowers the opacity a bit, and the
threshold is a bit higher, but the conclusions on the escape of $\g$
rays are not changed drastically.

Finally, I explore the ratio of the Compton-scattered flux to
synchrotron flux for a blob moving out in a jet.  This ratio, and
how it changes with time, can be a clue to finding the location of the
$\g$-ray emitting region.  For flares lasting longer than $\sim$\ a
day, the flaring region rapidly moves out of the BLR (for fiducial
values $\G=\delta_D=40$), so that scattering of BLR photons by a
single blob cannot explain these flares, although emission by many
separate blobs within the BLR would be possible.  Scattering of
photons from an extended dust torus at $\sim10\ \pc$ from the black
hole is not able to cool photons fast enough to explain the rapid
flares seen from FSRQs.  If $\g$-ray flares originate from EC at these
distances, another source of seed photons is needed to rapidly cool
the electrons producing these flares.  A possible alternative photon
source is other regions of the jet: such as a sheath surrounding the
emitting region \citep{ghisellini05,macdonald15,sikora16} or
scattering of photons from a standing shock or other moving regions of
the jet \citep{marscher14}.  The location of the $\g$-ray emitting
region in blazars and the seed photon source is still unclear.

\acknowledgements 

In Section \ref{signaturesection} I used data taken and assembled by
the WEBT collaboration and stored in the WEBT archive at the
Osservatorio Astrofisico di
Torino-INAF\footnote{\url{http://www.oato.inaf.it/blazars/webt/}}.  I
would particularly like to thank C.\ Raiteri for retrieving these
data.  This research has made use of the NASA/IPAC Extragalactic
Database (NED) which is operated by the Jet Propulsion Laboratory,
California Institute of Technology, under contract with the National
Aeronautics and Space Administration.  I am grateful to the
anonymous referee for a prompt and helpful report,} C.\ D.\ Dermer for
useful discussions on Compton scattering, and C.\ Done who
piqued my interest in beaming patterns for what became Section
\ref{beampatternsection}.  I am funded by the Chief of Naval Research.

\appendix

\section{Simple Empirical Stratified Broad Line Region Model}
\label{BLRmodel}

There is a long history of using composite quasar templates to
estimate the broad line luminosities and energy densities for blazars.
For instance, \citet{celotti97}; \citet{liu06}; and \citet{joshi14},
among others, used the template from \citet{francis91}; while
\citet{cerruti13_3c454} used the template from \citet{telfer02}.  When
performing $\gamma$-ray Compton scattering or $\g\g$ absorption
calculations, this has occasionally been criticized for not taking
into account the stratification of the broad line region, with
different lines being emitted at different radii
\citep[e.g.,][]{poutanen10,stern14}.  Here I present a method that
uses the composite quasar spectrum from the SDSS
\citep{vandenberk01} to estimate the radii where the lines are
primarily emitted, along with the lines' luminosities.

\begin{figure*}
\vspace{10.0mm} 
\epsscale{1.0} 
\plottwo{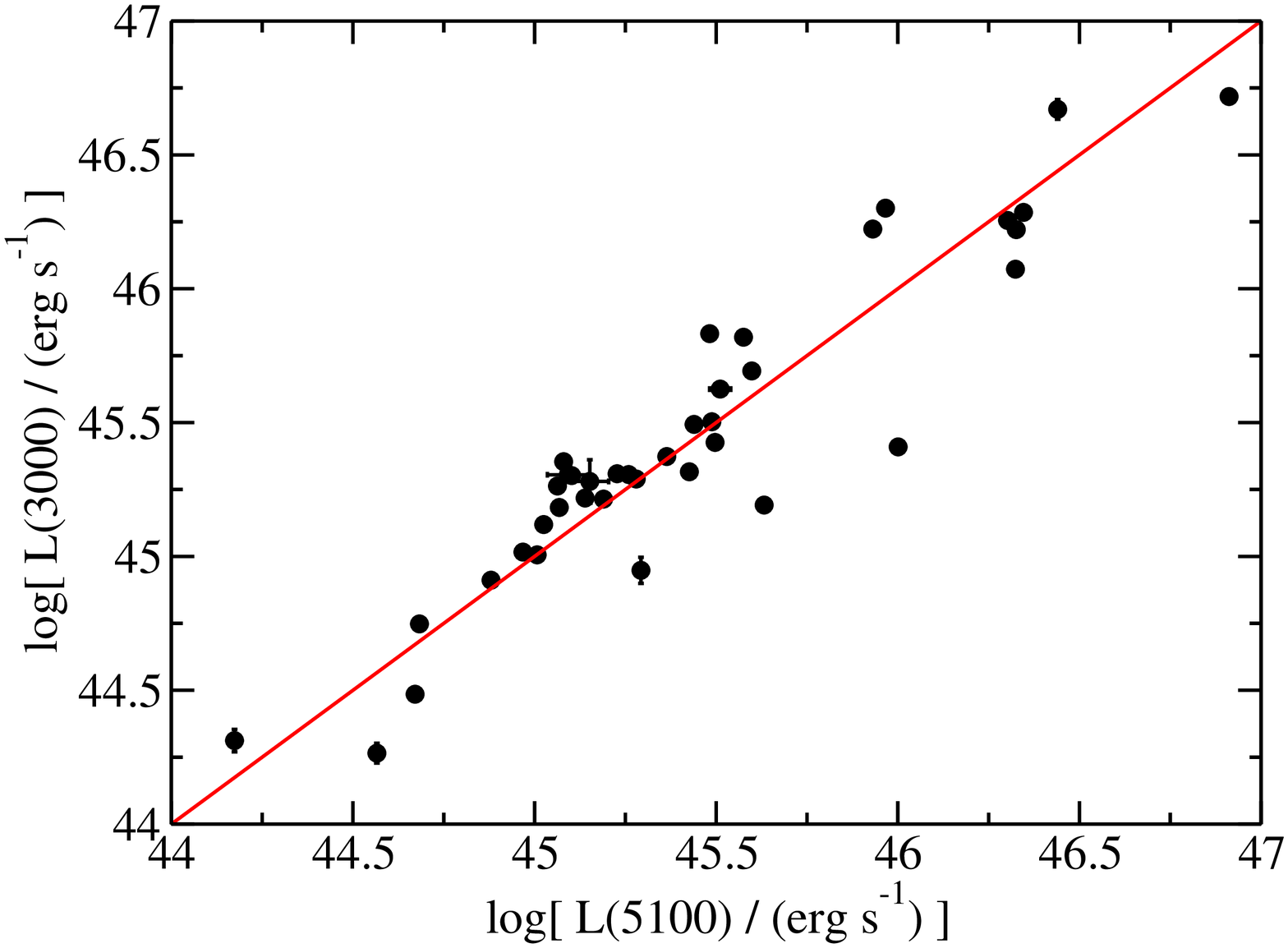}{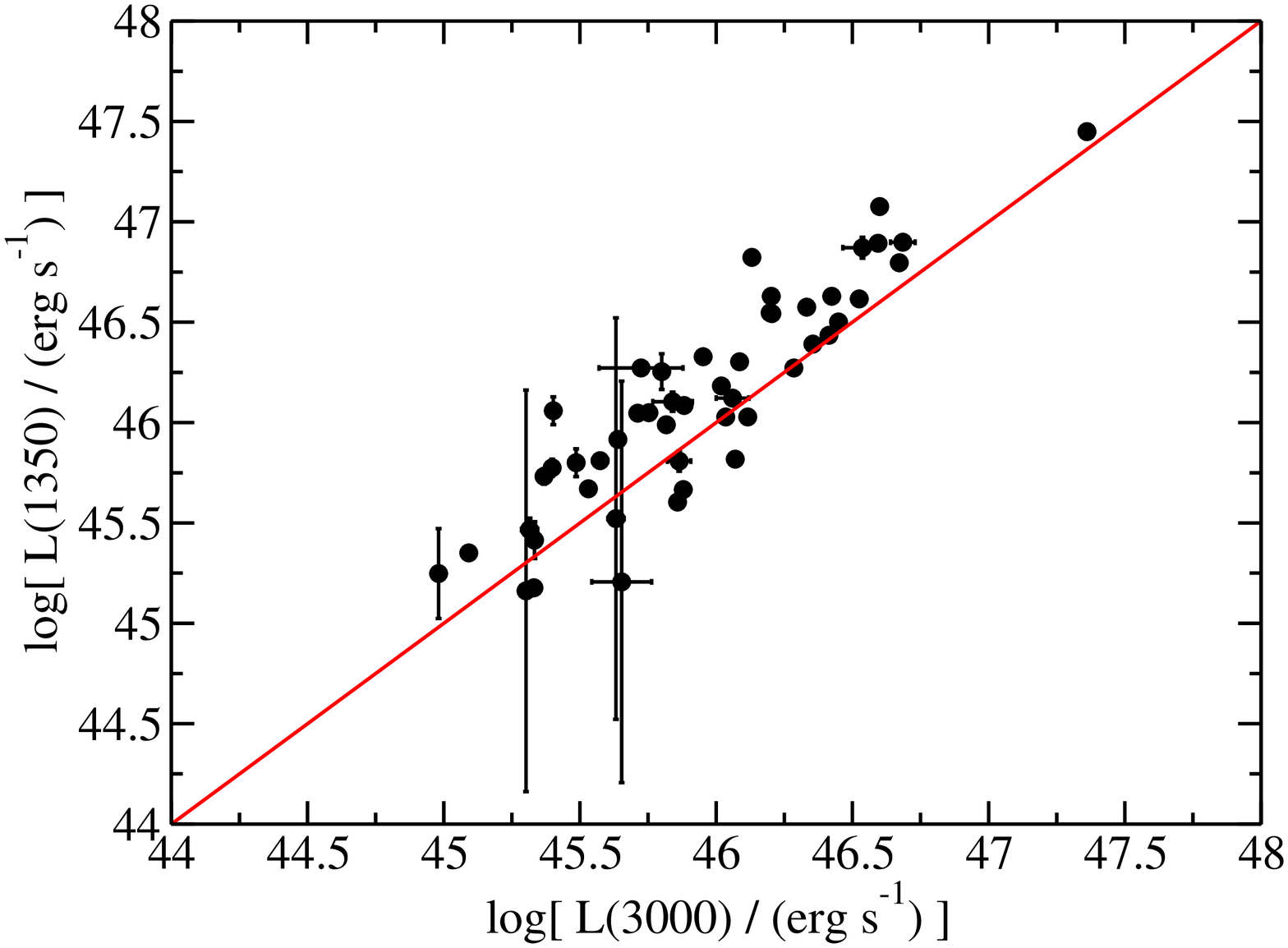}
\caption{Continuum luminosities of FSRQs at different wavelengths,
plotted against each other.  Data are from from \citet{shaw12}.  {\em
Left}: Luminosities at 3000 \AA\ versus 5100 \AA.  {\em Right}:
Luminosities at 1350 \AA\ versus 3000 \AA.  The red lines show where the
luminosities are equal.}
\label{Lcont1}
\vspace{2.2mm}
\end{figure*}

Assume that there are two quasars, $a$ and $b$, and each has two broad
emission lines seen in their spectra, lines 1 and 2.  Reverberation
mapping indicates that the continuum luminosity at a particular
wavelength (which can be different for each line) is proportional to
the square of the broad line radius
\citep[e.g.,][]{kaspi05,kaspi07,bentz06,bentz09,bentz13}, so that for
line 1,
\begin{flalign}
\label{La1}
\frac{L_{a,1}}{L_{b,1}} = \left( \frac{R_{a,1}}{R_{b,1}} \right)^2
\end{flalign}
and similarly for line 2, 
\begin{flalign}
\label{La2}
\frac{L_{a,2}}{L_{b,2}} = \left( \frac{R_{a,2}}{R_{b,2}} \right)^2 \ .
\end{flalign}
Reverberation mapping is usually done for radio quiet AGN, but a
handful of blazars have been the targets of reverberation mapping
campaigns as well, and they seem to obey the same $L\propto R^2$
relation as radio quiet AGN \citep[e.g.,][]{kaspi07}.  This is
explained by photoionization, with lines emitting at primarily one
radius where the ionization parameter and density are optimized
\citep[e.g.,][]{wandel97,wandel99}.  Reverberation mapping for several
objects also indicates that the broad line velocity width is related
to the broad line radius as one would expect if the broad line clouds
are in Keplerian orbits around the black hole
\citep{peterson99,peterson00,kollat03,peterson14}, so that
\begin{flalign}
\label{Va1}
\left( \frac{V_{a,1}}{V_{a,2}}\right)^2 = \frac{R_{a,2}}{R_{a,1}}
\end{flalign}
 and similarly for object $b$.  Combining Equations (\ref{La1})
through (\ref{Va1}) gives
\begin{flalign}
\label{Va2}
\left( \frac{V_{a,1}}{V_{a,2}}\right)^2 = \frac{R_{a,2}}{R_{a,1}} = 
\frac{R_{b,2}}{R_{b,1}}\left( \frac{L_{a,2}}{L_{a,1}}\right)^{1/2}
\left( \frac{L_{b,1}}{L_{b,2}}\right)^{1/2}
\end{flalign}
If the continuum luminosities at different wavelengths 
for a certain object are directly proportional to each other,i.e., if
$L_{a,1}\propto L_{a,2}$ and $L_{b,1}\propto L_{b,2}$, then 
\begin{flalign}
\label{Va3}
\left( \frac{V_{a,1}}{V_{a,2}}\right)^2 = 
\frac{R_{a,2}}{R_{a,1}} = 
\frac{R_{b,2}}{R_{b,1}} = 
\left( \frac{V_{b,1}}{V_{b,2}}\right)^2 
\end{flalign}
and the ratios of radii for different lines are proportional to each
other.  In Figure \ref{Lcont1} I plot the
optical/UV continua for FSRQs, taken from \citet{shaw12}, against each
other.  Not only does it seem that $L_{a,1}\propto L_{a,2}$, it
appears that $L_{a,1} \approx L_{a,2}$, as a first order
approximation, although the scatter is quite large.

Reverberation mapping is expensive in terms of telescope time; it has
only been performed for a limited number of objects and lines.  Still,
if one assumes that the empirical results represented by Equations
(\ref{La1}) through (\ref{Va3}) are correct for {\em all} broad lines
and quasars, one can use the composite quasar spectral template from
\citet{vandenberk01} to estimate the relative luminosities, line
velocity widths, radii, and energy densities for $r<R_{\rm li}$, 
\begin{flalign}
u = \frac{L_{\rm li}}{4\pi c R_{\rm li}^2}\ ,
\end{flalign}
for all lines in the template.  This is done in Table \ref{BLR_table}
 for all lines in the template [see Table 2 of \citet{vandenberk01}]
 with $V>1000\ \km\ \s^{-1}$.  All results are in terms of H$\beta$,
 the most common line used for reverberation mapping.  If one assumes
 the \citet{vandenberk01} template is representative of a quasar with
 a ``composite mass'', and the other assumptions stated above are
 correct, then these ratios should be approximately correct for {\em
 every} quasar.  This may also be justified by noting that composite
 SDSS quasar spectra, binned by luminosity, are very similar
 \citep{vandenberk04}.

How do the results compare with the results of reverberation mapping
campaigns?  Few campaigns have been performed on lines other than
H$\beta$, so comparisons are limited.  A comparison of
\ion{Si}{4}$\lambda1400$ and \ion{C}{4}$\lambda1549$ give comparable
radii, within a factor of 2.  The lines \ion{C}{3}]$\lambda1909$, and
\ion{He}{2}$\lambda4686$ fare worse, off by factors of about 3 and
$\ga80$, respectively.  A comparison of \ion{He}{1}$\lambda5876$ with
\citet{kollat03} also fares poorly, off by a factor of $\ga8$.  The
poor agreement may be because the luminosities for these lines are
more faint, making it difficult to measure their widths.  In this
case, it seems likely that the stronger lines in Table \ref{BLR_table}
are reliable within a factor of 2 or so, while the weaker ones are
completely unreliable.  Due to the poor agreement, I replaced the
results for \ion{He}{2}$\lambda4686$ and \ion{He}{1}$\lambda5876$ in
Table \ref{BLR_table} with those from \citet{peterson99} and
\citet{kollat03}.  The luminosities and velocity widths from Table
\ref{BLR_table} can also be compared with those found for FSRQs by,
for example, \citet{pian05} and \citet{shaw12}.  Although errors for
individual objects are large, agreement with the means of these
parameters is good, within about a factor of 2.

For NGC 5548, reverberation mapping indicates $R({\rm
Ly}\alpha)\approx 6$\ light-days \citep{derosa15} and $R({\rm
H}\beta)\approx 20$\ light-days \citep{peterson99}, a ratio in
approximate agreement with Table \ref{BLR_table} (0.30 versus 0.26).
This is perhaps the most reassuring, since Ly$\alpha$ is the brightest
line with the greatest energy density, due to its compactness and
brightness.  It is the most important line for the purposes of Compton
scattering and $\g\g$ absorption for $r<R({\rm Ly}\alpha)$.



As an example, one can consider the well-known $\g$-ray emitting
quasar 3C 454.3.  According to \citet{isler13}, the luminosity of
H$\beta$ is $L({\rm H}\beta) = (4.2\pm1.6)\times10^{43}\ \erg\
\s^{-1}$.  Using the relation from \citet{greene05},
\begin{flalign}
\left(\frac{L({\rm 5100\ \AA})}{10^{44}\ \erg\ \s^{-1}}\right) = \left(\frac{L({\rm
H}\beta)}{(1.425\pm0.007)\times10^{42}\ \erg\ \s^{-1}}\right)^{0.8826\pm0.0039}\ ,
\end{flalign}
one gets $L({\rm 5100\ \AA})=(2.0\pm0.7)\times10^{45}\ \erg\ \s^{-1}$ for the 
continuum luminosity at 5100 \AA.  Using the relation for the
radius where H$\beta$ is primarily emitted \citep{bentz13}, 
\begin{flalign}
R({\rm H}\beta) = 10^{16.94\pm0.03}\ \left( \frac{L({\rm 5100\
\AA})}{10^{44}\ \erg\ \s^{-1}} \right)^{0.533\pm0.035}\ \cm\ ,
\end{flalign}
one gets $R({\rm H}\beta)=(4.3\pm 0.2) \times10^{17}\ \cm$.  The energy density of 
H$\beta$ photons inside $R(\rm H\beta$) is then
\begin{flalign}
u({\rm H}\beta) = \frac{L({\rm H}\beta)}{4\pi c [R({\rm H}\beta)]^2} =
(6.0\pm2.4)\times10^{-4}\ \erg\ \cm^{-3}\ .
\end{flalign}
One can then estimate the luminosities, radii, and energy densities of
other broad lines using Table \ref{BLR_table}.

\begin{deluxetable}{llcccc}
\tabletypesize{\scriptsize}
\tablecaption{Broad Emission Line Parameters}
\tablewidth{0pt}
\tablehead{
\colhead{Ion} & 
\colhead{$\lambda$ [\AA]} &
\colhead{$L/L(H\beta)$} &
\colhead{$V/V(H\beta)$} &
\colhead{$R/R(H\beta)$} &
\colhead{$u/u(H\beta)$} 
}
\startdata
Ly$\epsilon$ &   937.80 & 0.24   & 0.61   & 2.7  &    0.033	\\
Ly$\delta$   &   949.74 & 0.24   & 0.60   & 2.8  &    0.031	\\
C III 	   &     977.02 & 0.60   & 1.1    & 0.83  &    0.88	\\
N III  	   &     990.69 & 0.60            & 1.1   & 0.85  &    0.83	\\
Ly$\beta$  &    1025.72 & 1.1             & 0.91   & 1.2  &    0.76	\\
O VI   	   &    1033.83 & 1.1             & 0.90   & 1.2  &    0.73	\\
Ar I   	   &    1066.66 & 0.094           & 0.47   & 4.5  &    0.0046	\\
Ly$\alpha$ &    1215.67 & 12              &  1.9  &  0.27 &    160	\\
O I    	   &    1304.35 & 0.23            & 0.50   & 4.0  &    0.014	\\
Si II      &    1306.82 & 0.23            & 0.50   & 4.0  &    0.014	\\
Si IV	   &    1396.76 & 1.0             & 1.1   & 0.83  &    1.5	\\
O IV]      &    1402.06 & 1.0             & 1.1   & 0.83  &    1.5	\\    
C IV	   &    1549.06 & 2.9             & 1.1   & 0.83  &    4.2	\\
N IV	   &    1718.55 & 0.030           & 0.51   & 3.8  &    0.0021	\\
Al II      &    1721.89 & 0.030           & 0.51   & 3.8  &    0.0021	\\
C III]     &    1908.73 & 1.8             & 1.5   & 0.46  &    8.9	\\ \relax
[Ne IV]    &    2423.83 & 0.051           & 0.42   & 5.8  &    0.0015	\\
Mg II      &    2798.75 & 1.7             & 1.5   & 0.45  &    8.6	\\
He I	   &    3188.67 & 0.051           & 0.48   & 4.3  &    0.003	\\
H$\delta$  &    4102.89 & 0.12            & 0.55   & 3.4  &    0.011	\\
H$\gamma$  &    4341.68 & 0.30            & 0.56   & 3.2  &    0.030	\\
He II\tablenotemark{a} &    4687.02 & 0.016           & 2.5    & 0.63 &    0.040    \\
H$\beta$ 	   &    4862.68 & 1.0     & 1.0   & 1.0  &     1.0	\\ \relax
[Cl III]   &    5539.43 & 0.039           & 0.46   & 4.8  &    0.0027	\\  
He I\tablenotemark{a}  &    5877.29 & 0.092           & 1.6   & 0.39  &    0.60	\\
H$\alpha$    &    6564.61 & 3.6           & 0.87   & 1.3  &    2.0	\\
\enddata
\tablenotetext{a}{Velocities and radii for these lines come from \citet{peterson99} and 
\citet{kollat03}.}
\label{BLR_table}
\end{deluxetable}


\bibliographystyle{apj}
\bibliography{variability_ref,EBL_ref,references,mypapers_ref,blazar_ref,sequence_ref,SSC_ref,LAT_ref,3c454.3_ref}

\end{document}